\documentclass{article} 
\usepackage{iclr2027_conference,times}

\usepackage{amsmath,amsfonts,bm}

\def\eqref#1{equation~\ref{#1}}

\def\1{\bm{1}}

\DeclareMathAlphabet{\mathsfit}{\encodingdefault}{\sfdefault}{m}{sl}
\SetMathAlphabet{\mathsfit}{bold}{\encodingdefault}{\sfdefault}{bx}{n}

\usepackage{hyperref}
\usepackage{url}
\usepackage{verbatim}
\usepackage{booktabs}
\usepackage{multirow}
\usepackage{booktabs}
\usepackage{graphicx}
\usepackage[table]{xcolor}
\usepackage{float}
\usepackage{adjustbox}
\usepackage{xspace}
\usepackage{threeparttable}
\usepackage{wrapfig}

\newcommand{\app}{\textsc{ARIA}\xspace}

\title{Ask Your Advisor: Learning When to\\
Consult Stronger Models in Agentic Systems}

\author{%
\bfseries Pascal Sowietzki, Steffen Holter, Yannick Metz, Emil Kasper, Raphael Baur,\\
\bfseries Mennatallah El-Assady\\
\normalfont ETH Zurich, Switzerland\\
\normalfont\footnotesize\texttt{\{psowietzki, kaspere\}@ethz.ch}\\
\normalfont\footnotesize\texttt{\{steffen.holter, yannick.metz\}@inf.ethz.ch}\\
\normalfont\footnotesize\texttt{\{raphael.baur, menna.elassady\}@ai.ethz.ch}%
}

\iclrfinalcopy

\begin{document}

\maketitle
\lhead{}

%

\newcommand{\fix}{\marginpar{FIX}}
\newcommand{\new}{\marginpar{NEW}}

\iclrfinalcopy 

\maketitle

\begin{abstract} 
To balance task performance and computational cost, agentic workflows can use a lower-cost model for execution and selectively consult a stronger model. In such executor–advisor systems, the executor decides when to seek advice, making escalation depend on self-assessment, although current language models struggle to detect where their own reasoning fails. The need for stronger reasoning emerges during execution and varies across workflow phases. We introduce \app, a meta-controller that learns when to consult by balancing task success against consultation cost. Our approach formulates consultation as a sequential resource-allocation problem. At workflow contact points, \app maps execution evidence to a compact state abstraction and selects \textsc{Silent} or \textsc{Consult}, with action values learned by Monte Carlo control from cost-penalized outcomes. Across evaluated tasks of SWE-bench Verified, Terminal-Bench 2.1, and $\tau^2$-bench, \app recovers 44--81\% of the executor-advisor success gap at less than half the advisor model's cost. \app achieves the best observed success-cost trade-off among ten other evaluated conditions, including routing and cascading methods, fixed and prompted consultation, and random placement at matched advice volume. The policies are compact and inspectable, with a single consultation-price parameter governing the success–cost trade-off without retraining either underlying model.

\end{abstract}

\section{Introduction}
Agentic systems increasingly tackle complex tasks such as software engineering~\citep{openai_2024_sweverified}, data science~\citep{merrill_2026_terminalbench}, and customer support~\citep{barres_2025_tau2bench} through long-running workflows that extend beyond isolated queries~\citep{openai_2025_agents,anthropic_2024_effectiveagents,mohan_2026_agentic}. 
Sustaining these workflows over long task horizons is costly, as each stage may require reasoning, tool use, and revision. However, prior work shows substantial scope for reducing inference cost while maintaining similar task performance ~\citep{kapoor_2025_agents}, and the need for stronger reasoning varies throughout execution. The challenge is therefore to allocate capability selectively, using stronger reasoning where its expected benefit justifies its additional cost.

Existing capability-allocation methods are selective, but mostly operate at the model or task level, making allocation decisions from the input before execution. Specialization \emph{changes} the executor \citep{wang_2024_codeact,belcak_2025_slm}, routing determines \emph{which model} acts \citep{aggarwal_2024_automix,ong_2025_routellm}, and adaptive computation controls \emph{how much reasoning} it receives before execution \citep{snell_2025_scaling,damani_2025_thinking}. These approaches reduce reliance on costly frontier-model capability, but do not allocate stronger reasoning to specific moments within an evolving workflow. In long-running execution, the value of stronger reasoning can change as new observations, errors, or recovery opportunities emerge. A routine task may contain only a few decision points where additional capability is particularly valuable.

Executor-advisor systems provide a natural mechanism for introducing stronger reasoning into a workflow. A lower-cost model remains responsible for execution, while a stronger advisor provides guidance without taking control \citep{anthropic_2026_advisor}. This allows routine steps to remain with the executor and advisor input to be added selectively as the workflow unfolds. The remaining question is \emph{when} to consult. Current approaches leave this to the executor itself, making escalation depend on self-assessment. Previous work has shown that large language models (LLMs) struggle to localize their own errors and have poorly calibrated confidence \citep{tyen_2024_errors,xiong_2024_uncertainty}. Uncertainty alone does not indicate whether consultation is useful. An
executor may recover, while advice may be valuable even when it appears confident. The relevant question is whether consulting at this point improves the final outcome enough to justify its cost. This separates consultation allocation from a complementary problem: whether the executor incorporates the advice it receives.

\begin{figure}
        \centering
        \includegraphics[width=1\linewidth]{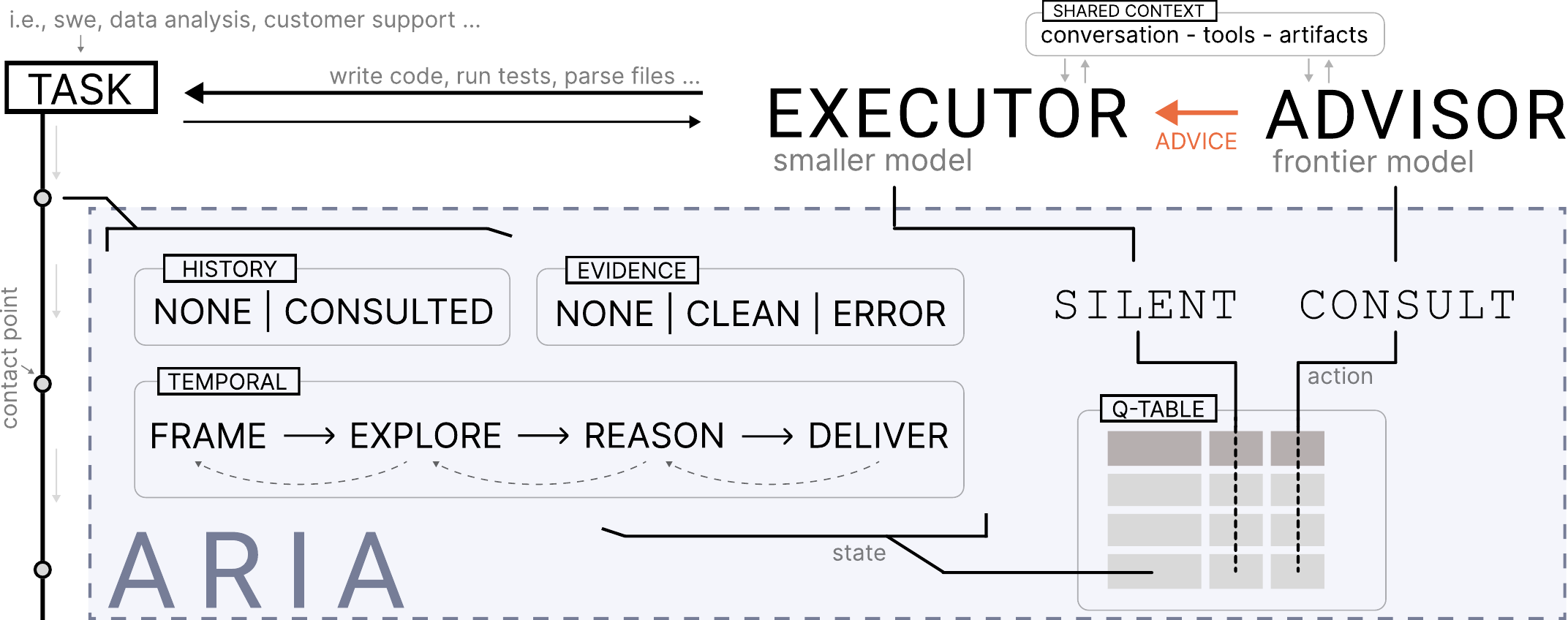}
        \vspace{-1.5em}
        \caption{
    \app setup. At each contact point, the meta controller chooses between \textsc{Silent} and \textsc{Consult}. Consultation adds guidance from a stronger advisor to the shared context.
}
        \label{fig:ARIA_setup}
    \vspace{-1em}
    \end{figure}

We introduce \textbf{\app} (Allocation of Reasoning In Agentic systems), a learned meta-controller for selective expert consultation in agentic workflows. \app regulates access to a stronger advisor while keeping a lower-cost executor in control. We instantiate \app on Magentic-One~\citep{fourney_2024_magenticone}, while the controller itself only assumes recurring decision points: At workflow decision points, such as replanning or final delivery, it chooses whether the executor continues or receives guidance. We formulate this policy as a sequential decision problem and learn it with tabular Monte Carlo control. The state is derived from observable workflow evidence rather than self-reported confidence, and neither language model is fine-tuned. This way, \app can learn an optimal trade-off between two deployment extremes by allocating advisor to selected points in the trajectory.

We evaluate \app across three agentic settings: SWE-bench Verified \citep{openai_2024_sweverified}, Terminal-Bench 2.1 \citep{merrill_2026_terminalbench}, and $\tau^2$-bench \citep{barres_2025_tau2bench}, spanning homogeneous, heterogeneous, and interactive workflows across software engineering, data analysis, system administration and customer service. We anchor the comparison with \textit{Executor-only} and \textit{Advisor-only} baselines as the lower and stronger-model reference points, and compare fixed, prompted, learned and allocation methods from related work between these reference points. A \textit{Random}-placement control approximates \app's consultation rate to separate the effect of consultation timing from simply having advisor access. We report task success with realized model cost and evaluate how much of the executor-advisor performance gap selective consultation can recover.

Our contributions are threefold. First, we introduce \app, a modular and interpretable framework for learning consultation policies over executor–advisor pairs. Second, we evaluate \app across three agentic domains, analyzing success, cost and efficiency. Third, we analyze failures in executor–advisor collaboration to gain further insights into inter-agent coordination. We release the framework and traces for research on capability allocation and agent collaboration.

\section{Related Work} 
\textbf{Specialized models and adaptive computation.}
Specialization and adaptive computation improve efficiency without conditioning on evidence revealed during the running workflow. Small language models can be trained or fine-tuned for recurring tasks \citep{belcak_2025_slm}. \textit{CodeAct} fine-tunes models on agent-environment interactions, while reasoning models specialize for mathematical and code reasoning through supervised fine-tuning and reinforcement learning \citep{wang_2024_codeact,liu_2026_acereason}. Inference-time methods instead allocate additional computation to improve reasoning \citep{snell_2025_scaling}. Input-adaptive allocation predicts the benefit of computation budgets from the query alone and assigns resources before generating a response \citep{damani_2025_thinking}. \app differs by allocating capability after execution has begun, using evidence revealed along the trajectory.

\textbf{Routing, cascading, and orchestration.}
Routing and orchestration allocate capability dynamically, but typically do so by changing which model executes or how execution is structured. \textit{RouteLLM} selects between stronger and weaker models \citep{ong_2025_routellm}, while \textit{AutoMix} uses self-verification and a POMDP cascade to decide whether a smaller model's answer should be accepted or escalated \citep{aggarwal_2024_automix}. Allocation can operate at finer granularity: STEER routes reasoning steps between smaller and larger models using logit-derived confidence estimates, requiring access to model-internal signals unavailable in our setup \citep{lee_2026_steer}. At the system level, \textit{AFlow} optimizes workflow structures, while \textit{Automated Design of Agentic Systems} searches over agent architectures from execution feedback \citep{zhang_2025_aflow,hu_2025_adas}. \app instead takes the executor and workflow as given and learns when to acquire guidance without model-internal signals.

\textbf{External consultation.}
Existing approaches make advice available during execution, but typically leave the decision \emph{when} to request it to the agent itself. Action-advising work treats guidance as a limited resource and shows that its effect can depend on when a fixed budget is spent \citep{torrey_2013_teaching}. In these settings, however, advice typically serves to accelerate the learning of an agent's task policy. These concepts are also emerging in human-agent systems that provide several interaction patterns: \textit{Magentic-UI} supports co-planning and human intervention during execution \citep{mozannar_2025_magenticui}, while \textit{Collaborative Gym} incorporates communication into the agent's action space and studies prompted situational planning \citep{shao_2026_collaborativegym}. Related ideas have recently appeared in agent-agent collaboration. Anthropic's Advisor Strategy pairs a lower-cost executor with a stronger advisor that can be consulted during execution through a tool call \citep{anthropic_2026_advisor}. \textit{SWE-Protégé} goes further by training a smaller software-engineering agent to recognize when expert help is needed and act on the resulting guidance \citep{kon_2026_sweprotege}, whereas \app keeps the models fixed. Our contribution is therefore not the introduction of advice itself. \app instead learns a cost-aware policy for when to consult the advisor during execution, without modifying the underlying model.

\section{Methodology} 
\label{sec:methodology}
\app keeps the executor, advisor, environment and workflow fixed and learns only a separate binary policy that explicitly determines when advisor guidance is allocated during execution.

\subsection{Consultation during the trajectory}
\label{sec:placing_decision}
Agentic workflows often involve recurring points at which agents must decide whether to continue, replan, or terminate in response to execution feedback \citep{agashe_2025_agents,huang_2023_inner}. We introduce opportunities for consultation at these decision points. Using Magentic-One as our execution framework \citep{fourney_2024_magenticone}, we define three orchestration events as consultation contact points: \textsc{Pre-Step}, before the next executor instruction; \textsc{Stall}, when stalled progress triggers replanning; and \textsc{Pre-Delivery}, immediately before final delivery. The contact-point type determines why a consultation opportunity occurs, whereas the temporal state in Section~3.2 separately captures coarse progress over the sequence of contact points.

At each contact point $t$, \app selects $a_t \in \{\mathrm{SILENT},\mathrm{CONSULT}\}$. Under \textsc{Silent}, the executor remains in control. Under \textsc{Consult}, the advisor receives task specification and current trajectory information and returns a natural-language assessment with recommended next steps. The orchestrator incorporates this guidance into the executor's subsequent instruction, while the executor remains responsible for reasoning, tool use, and final submission. Consultation can affect later actions, observations, and contact points through the executor's response to advice; exact intervention semantics are given in Appendix~\ref{app:advisor-semantics}. \app learns consultation value from downstream outcomes rather than immediate local feedback.

\subsection{State abstraction}
\label{sec:states}
\app conditions its decision on existing signals available from the workflow rather than on the executor's own assessment, motivated by known limitations of LLM self-diagnosis and confidence calibration \citep{tyen_2024_errors,xiong_2024_uncertainty}. We summarize the recorded history $H_t$ along three dimensions: temporal position $p_t$ (4 categories), execution evidence $e_t$ (3), and consultation history $h_t$ (2),

\[
    s_t=\phi(H_t)=(p_t,e_t,h_t),
    \qquad
    |\mathcal{S}|=4\times3\times2=24.
\]
Each dimension distinguishes circumstances under which the value of consultation may differ.

The contextual dimension captures what execution has revealed. Execution feedback can support correction \citep{chen_2024_selfdebug, shinn_2023_reflexion}, while successful self-repair depends strongly on the feedback available \citep{olausson_2024_repair, huang_2024_selfcorrect}. We therefore use $e_t\in\{\textsc{None},\textsc{Clean},\textsc{Error}\}$ to distinguish absent evidence from output without or with a recognized error signal. This reflects observed execution status rather than correctness: clean output may follow a wrong strategy, and an error may be recoverable unaided. The temporal dimension captures where this evidence occurs and is derived independently of the contact-point type. \textsc{Pre-Delivery} maps directly to \textsc{Deliver}; otherwise, the first contact point maps to \textsc{Frame}, the next two to \textsc{Explore}, and later contact points to \textsc{Reason}. Inspired by \citet{polya_1945_solve}, $p_t\in\{\textsc{Frame},\textsc{Explore},\textsc{Reason},\textsc{Deliver}\}$ provides a task-agnostic representation of coarse workflow progress. The consultation-history dimension, $h_t\in\{0,1\}$, records whether advisor guidance has already entered the trajectory. Because prior consultation changes the information available to the executor, the value of a further request may differ, a distinction also reflected in contemporary executor–advisor systems \citep{anthropic_2026_advisor}. Remaining budget is omitted from $s_t$; once $B$ is exhausted, \textsc{Consult} is forced to \textsc{Silent}. Thus, the policy does not condition explicitly on remaining budget. The 24-state abstraction keeps the policy compact and inspectable by summarizing the trajectory into a small number of workflow signals; Appendix~\ref{app:state_construction} specifies the exact state mapping.\looseness=-1

\subsection{Cost-aware policy learning}
\label{sec:learn}
The state representation provides no intermediate signal of consultation value. \app therefore learns from the terminal task outcome. For trajectory $\tau$, let $Y(\tau)\in\{0,1\}$ denote task success and $N_c(\tau)$ the number of consultations. We define the cost-aware return
\begin{equation}
    G(\tau)
    =
    Y(\tau)
    -
    \lambda\,\frac{\bar{c}}{\Delta}\,N_c(\tau),
    \label{eq:aria-return}
\end{equation}
where $\bar{c}$ is the mean consultation cost and $\Delta$ the mean per-episode spend of consulting at every contact point, measured per domain (Appendix~\ref{app:training}). Thus, $\bar{c}/\Delta$ normalizes one consultation against the spend of consulting throughout, making $\lambda$ approximately the price of that policy in units of one solved task. We assign a fixed penalty to each consultation, treating advisor access as a count-limited resource, consistent with prior work on advice budgets and constant intervention costs \citep{torrey_2013_teaching,cortes_2016_learning}. Maximizing the expected return in~\eqref{eq:aria-return} yields a fixed-multiplier Lagrangian relaxation of maximizing expected task success subject to a consultation-cost budget \citep{altman_1999_constrained}, with $\kappa=\lambda\bar{c}/\Delta$ as the effective per-consultation penalty. LLM collaboration systems similarly use a fixed deferral penalty to control escalation behavior \citep{yang_2026_adaptive}.

Because the task outcome arrives only at the end of the episode, \app learns action values using every-visit tabular Monte Carlo control with sample-average updates \citep{sutton_2018_reinforcement}. A visit at contact point $t$ is assigned the remaining episodic return
$G_t = Y(\tau) - \kappa\,N_c^{\geq t}(\tau)$, the task outcome minus
penalties for consultations at or after $t$; the episodic return in~\eqref{eq:aria-return} is $G_0$. We set $\gamma=1$: discounting would reduce the terminal outcome relative to immediate consultation cost and bias the policy toward later consultation. Training uses an $\epsilon$-greedy behavior policy, and only episodes reaching a terminal outcome update the table. Evaluation uses the frozen greedy policy, with ties, including unvisited states, resolving to \textsc{Silent}. Further details are reported in Appendix~\ref{app:training}.

\section{Experiments}
\label{sec:setup}
We evaluate \app across three agentic environments using fixed proprietary models. Direct baselines are limited to methods that do not require model internals or fine-tuning. We test whether learned consultation improves the success–cost trade-off over autonomous and task-level baselines, whether placement matters beyond advisor access, how consultation shifts the operating point, and what the results reveal about agent-agent collaboration.

\textbf{Benchmarks and models.}
We evaluate software engineering on SWE-bench Verified \citep{openai_2024_sweverified}, heterogeneous terminal workflows on Terminal-Bench~2.1 \citep{merrill_2026_terminalbench}, and interactive customer service on $\tau^2$-bench \citep{barres_2025_tau2bench}. Train and test tasks are disjoint. We use 50/50 train/test tasks for SWE-bench, with Verified-Mini as the test set, 27/50 for Terminal-Bench and 20/40 tasks for $\tau^2$-bench. Appendix~\ref{app:splits} details all benchmark splits. The executor is \texttt{gemini-3-flash-preview} at low reasoning effort and the advisor is \texttt{gemini-3.1-pro-preview} at high reasoning effort. Using models from the same family reduces cross-model differences in deployment behavior. Consultation conditions use at most $B=4$ advisor calls per episode; budget sensitivity is examined in Section~\ref{sec:placement}.

\textbf{Evaluation conditions.}
Our baselines are chosen to test competing explanations for the benefit of learned consultation. \textit{Executor-only} and \textit{Advisor-only} establish the capability range. \textit{FrugalGPT} \citep{chen_2024_frugalgpt}, \textit{RouteLLM} \citep{ong_2025_routellm}, and \textit{AutoMix} \citep{aggarwal_2024_automix} test whether routing or cascading provides sufficient allocation against \app under the same model pair. \textit{Plan}, motivated by Magentic-UI's upfront workflow planning \citep{mozannar_2025_magenticui}, \textit{Heuristic}, derived from Anthropic's recommended start/end and stall triggers \citep{anthropic_2026_advisor}, and \textit{Always} test whether defined consultation schedules or advisor access without learned timing are sufficient. \textit{LLM-Router} tests whether the native LLM orchestrator can decide when to involve the advisor without a learned controller, while \textit{Random} separates informed timing from consultation volume. Together, these baselines isolate the roles of model strength, allocation strategy, advisor access, and consultation timing. Appendices~\ref{app:baselines} and~\ref{app:random-calibration}
provide baseline details and \textit{Random}-control calibration.

\vspace{-1em}
\begin{table}[H]
\centering
\footnotesize
\setlength{\tabcolsep}{5pt}
\caption{Baseline design for selective consultation, covering capability references, task-level allocation, fixed schedules, LLM-based timing, uninformed placement, and learned allocation.}
\label{tab:conditions}
\begin{tabular}{@{}lll@{}}
\toprule
Condition & Allocation rule & Tests for \\
\midrule
Executor-only
    & no advisor
    & lower-cost reference \\

Advisor-only
    & advisor executes
    & stronger-model reference \\

\addlinespace
FrugalGPT {\scriptsize\citep{chen_2024_frugalgpt}}
    & score-based cascade
    & cost-aware task-level cascading \\

RouteLLM {\scriptsize\citep{ong_2025_routellm}}
    & pre-task routing
    & learned task-level routing \\

AutoMix {\scriptsize\citep{aggarwal_2024_automix}}
    & verify then escalate
    & adaptive task-level escalation \\

\addlinespace
Plan {\scriptsize\citep{mozannar_2025_magenticui}}
    & first contact
    & early intervention \\

Heuristic {\scriptsize\citep{anthropic_2026_advisor}}
    & beginning/end triggers
    & vendor-reported heuristic \\

Always
    & at every contact point
    & advisor access without selection \\

LLM-Router
    & orchestrator agent selects advisor
    & LLM-decided/prompted consultation\\

\addlinespace
Random
    & rate-matched advice
    & matched consultation volume \\

\addlinespace
\app
    & state-conditioned policy
    & learned consultation timing \\
\bottomrule
\end{tabular}
\end{table}
\vspace{-1,7em}

\textbf{Consultation-price ablation.} The default policy ($\lambda=0.10$) is trained online under its own $\epsilon$-greedy behavior policy. Policies for $\lambda \in \{0.35,0.50\}$ are re-estimated offline from the same training trajectories by recomputing returns and the value table. Because the trajectories were collected under the $\lambda=0.10$ behavior policy, this sweep is an offline policy re-estimation under that visitation distribution rather than equivalent retraining at each price (Appendix~\ref{app:training}).

\textbf{Quantitative evaluation.}
Following \textit{RouteLLM} \citep{ong_2025_routellm}, we report \emph{performance gap recovered} (PGR), which measures the fraction of the executor–advisor success gap recovered by a condition. Because the executor remains the base model across evaluated baselines, consultation adds advisor computation to the \textit{Executor-only} costs. We therefore normalize realized evaluation cost between the \textit{Executor-only} (0) and \textit{Advisor-only} (1) anchors:
\[
    \mathrm{PGR}
    =
    \frac{S-S_{\mathrm{exec}}}
         {S_{\mathrm{front}}-S_{\mathrm{exec}}},
    \qquad
    \mathrm{CostShare}
    =
    \frac{C-C_{\mathrm{exec}}}
         {C_{\mathrm{front}}-C_{\mathrm{exec}}}.
\]
PGR is normalized by the same anchors, such that \textit{Executor-only} maps to 0 and \textit{Advisor-only} to 1 on both axes. A task-independent mixture lies on $\mathrm{PGR}=\mathrm{CostShare}$ and thus has $\mathrm{ER}=0$; positive values indicate a better success-cost trade-off than simple model mixing. We define \emph{excess recovery} as $\mathrm{ER}=\mathrm{PGR}-\mathrm{CostShare}$, the vertical distance from this reference line. Positive ER therefore means that a method recovers more performance than its normalized cost share, with larger values indicating a better success-cost trade-off. Detailed token and cost reconciliation are reported in
Appendix~\ref{app:tokens}.

\textbf{Qualitative failure analysis.}
To analyze which failures remain beyond success rates and costs, we apply the Multi-Agent System Failure Taxonomy (MAST) \citep{cemri_2025_multiagent}, which groups 14 failure modes into system-design (FC1), inter-agent misalignment (FC2), and task-verification (FC3). Each failed trajectory receives one primary failure mode and additionally, all modes that occur. MAST annotations are produced by an LLM judge and are therefore validated on a stratified subset against human annotations; the full setup and protocols are documented in Appendix~\ref{app:mast}.

\section{Results}
\label{sec:results}
On SWE-bench, performance ranges from 18.0\% at \$11.48 for \textit{Executor-only} to 62.0\% at \$45.90 for \textit{Advisor-only}, with \app reaching 48.0\% at \$22.09 (Table~\ref{tab:main}). The same pattern holds across the other benchmarks: \app achieves a performance gap recovered (PGR) of 68.2\%, 44.4\%, and 80.6\% on SWE-bench, Terminal Bench, and $\tau^2$-bench while using 48.1\%, 40.9\%, and 34.5\% of \textit{Advisor-only} evaluation cost. Its excess recovery ($\mathrm{ER}=\mathrm{PGR}-\mathrm{CostShare}$), measuring how much more performance gap is recovered than normalized cost is spent, is 0.37, 0.15, and 0.65, respectively, exceeding all other evaluated conditions. Moreover, no evaluated condition achieves higher mean success than \app at equal or lower cost, or lower cost at equal or higher mean success. Thus, \app lies on the empirical success–cost Pareto frontier among the evaluated conditions.

\begin{table}[H]
\centering
\footnotesize
\setlength{\tabcolsep}{4pt}

\caption{
Success-cost trade-offs over three seeds (standard deviations are reported in Appendix~\ref{app:dispersion})
}
\label{tab:main}

\begin{adjustbox}{max width=\linewidth}
\begin{tabular}{@{}l rrr|rrr|rrr@{}}
\toprule
& \multicolumn{3}{c|}{\textbf{SWE-bench Verified}}
& \multicolumn{3}{c|}{\textbf{Terminal-Bench 2.1}}
& \multicolumn{3}{c}{\textbf{$\tau^2$-bench Telecom}} \\
\cmidrule(lr){2-4}
\cmidrule(lr){5-7}
\cmidrule(lr){8-10}

Condition
& Succ.\ \% & Cost \$ & ER
& Succ.\ \% & Cost \$ & ER
& Succ.\ \% & Cost \$ & ER \\
\midrule

Executor-only
& 18.0 & 11.48 & 0.00
& 21.3 & 9.76 & 0.00
& 72.5 & 5.07 & 0.00 \\

Advisor-only
& 62.0 & 45.90 & 0.00
& 51.3 & 60.99 & 0.00
& 98.3 & 22.50 & 0.00 \\

\addlinespace
FrugalGPT
& 31.3 & 22.61 & -0.02
& 22.0 & 11.10 & 0.00
& 78.3 & 7.02 & 0.11 \\

RouteLLM
& 30.0 & 25.03 & -0.12
& 28.0 & 21.66 & -0.01
& 88.3 & 9.61 & 0.35 \\

AutoMix
& \textbf{52.0} & 59.68 & -0.63
& 22.0 & 16.51 & -0.11
& \textbf{97.5} & 24.98 & -0.17 \\

\addlinespace
Plan
& 24.0 & 10.90 & 0.15
& 24.0 & 11.34 & 0.06
& 70.0 & 5.69 & -0.13 \\

Heuristic
& 41.3 & 21.50 & 0.24
& \textbf{34.7} & 27.47 & 0.10
& 83.3 & 7.13 & 0.30 \\

Always
& 34.0 & 21.39 & 0.08
& \textbf{35.3} & 27.94 & 0.11
& 87.5 & 9.01 & 0.36 \\

LLM-Router
& 29.3 & 15.77 & 0.13
& \underline{30.0} & 18.47 & 0.12
& 87.5 & 6.16 & 0.52 \\

Random
& 28.0 & 17.33 & 0.06
& \underline{30.7} & 22.25 & 0.07
& 85.8 & 8.24 & 0.33 \\

\midrule
\textbf{\app}
& \underline{48.0} & 22.09 & \textbf{0.37}
& \textbf{34.7} & 24.93 & \textbf{0.15}
& \underline{93.3} & 7.77 & \textbf{0.65} \\

\bottomrule
\end{tabular}
\end{adjustbox}
\end{table}

\textbf{Learned consultation differs from task-level allocation.}
The task-level allocation baselines do not reproduce the same success-cost position as our approach. On SWE-bench, \textit{FrugalGPT}, \textit{RouteLLM}, and \textit{AutoMix} achieve ER values of $-0.02$, $-0.12$, and $-0.63$, compared with 0.37 for \app. \textit{RouteLLM} and \app also spend similar amounts (\$25.03 vs.\ \$22.09), yet solve 30.0\% and 48.0\% of tasks. The pattern is similar on Terminal-Bench, where all three task-level methods remain at or below the executor–advisor reference line while \app reaches ER $=0.15$. On $\tau^2$-bench, \textit{FrugalGPT} and \textit{RouteLLM} improve over the reference line but remain below \app's ER of 0.65; \textit{AutoMix} reaches 97.5\% at an evaluation cost of \$24.98, 3.2$\times$ that of \app and higher than even \textit{Advisor-only} execution. The evaluated allocation methods do not match \app's observed success-cost trade-off, suggesting value in learning allocation on evidence emerging during execution. \textit{FrugalGPT} and \textit{RouteLLM} use task-level information, while \textit{AutoMix} additionally uses post-generation self-verification. None of these approaches incorporates evolving evidence throughout the agentic trajectory, specifically on the tested benchmarks.

\begin{wraptable}[9]{r}{0.51\textwidth}
\vspace{-1.8\baselineskip}
\centering
\footnotesize
\setlength{\tabcolsep}{4pt}

\caption{\app relative to the autonomous anchors.}
\label{tab:aria-relative}

\begin{tabular}{@{}lccc@{}}
\toprule
& \textbf{SWE}
& \textbf{Terminal}
& \textbf{$\tau^2$} \\
\midrule
PGR
& 68.2\% & 44.4\% & 80.6\% \\
Cost / Pro
& 48.1\% & 40.9\% & 34.5\% \\
Cost Share
& 30.8\% & 29.6\% & 15.5\% \\
Sav. vs.\ equal-succ. mix
& \$12.86 & \$7.59 & \$11.36 \\
\bottomrule
\end{tabular}

\vspace{1pt}
{\scriptsize
PGR follows the weak-strong performance-gap normalization of\\ \citet{ong_2025_routellm}. ER is defined as $\mathrm{PGR}-\mathrm{CostShare}$.}

\vspace{-0.4\baselineskip}
\end{wraptable}

\textbf{Advisor access alone does not explain the result.} The \textit{Always} condition consults at every opportunity under the same budget as \app. On SWE-bench it costs nearly the same (\$21.39 vs.\ \$22.09) but achieves 14.0 pp less; on $\tau^2$-bench it is both more expensive and 5.8 points less successful. Terminal-Bench provides the weaker contrast as \textit{Always} gains 0.6 points in success but costs 12\% more.\looseness=-1

\subsection{Placement matters beyond advisor access and volume}
\label{sec:placement}

\textbf{Similar advice volume produces different outcomes.}
The aggregate comparisons do not yet separate consultation placement from consultation volume. Similar advice volume produces different outcomes: On SWE-bench, \textit{Random}, \textit{Heuristic}, and \app use 2.53, 2.75, and 2.78 consultations per episode, yet achieve 28.0\%, 41.3\%, and 48.0\% success, respectively (Appendix~\ref{app:placements}). \app gains 20.0 pp over uninformed placement at similar advice volume, with a smaller 6.7-point advantage over \textit{Heuristic}. The effect is weaker on Terminal-Bench. \textit{Random} reaches 30.7\% success with 2.42 consultations per episode, while \textit{Heuristic} and \app both reach 34.7\% with 2.25 and 2.46 consultations; \app costs 9.2\% less than \textit{Heuristic}. On $\tau^2$-bench, \app reaches 93.3\% success versus 85.8\% for \textit{Random} and is slightly cheaper, but \textit{Random} realizes only 87.8\% of \app's consultation volume. The placement control is tightest on SWE-bench and weaker on $\tau^2$.\looseness=-1

\begin{wraptable}[7]{r}{0.29\linewidth}
\vspace{-0.8\baselineskip}
\centering

\begin{threeparttable}
\caption{Change from \textit{Always-$B4$} to \textit{$B10$}.}
\label{tab:always-budget}

\scriptsize
\setlength{\tabcolsep}{2.5pt}
\renewcommand{\arraystretch}{0.92}

\begin{tabular}{@{}lrrrr@{}}
\toprule
& $\Delta S$
& $\Delta C$
& $\Delta N_c$
& $\Delta$ER \\
\midrule
SWE
& +20.7
& +\$6.42
& +1.95
& +0.28 \\
TB
& +4.0
& +\$6.74
& +0.82
& 0.00 \\
$\tau^2$
& -2.5
& +\$5.04
& +5.78
& -0.38 \\
\bottomrule
\end{tabular}

\vspace{-0.6\baselineskip}
\end{threeparttable}
\end{wraptable}

\textbf{More advice is not uniformly better.} In the headline comparison, \textit{Always} uses the same $B=4$ consultation budget as \app{} in order to have equal conditions. Even under this matched budget, it falls short of \app{} by 14.0 points on SWE-bench and 5.8 points on $\tau^2$-bench; on Terminal-Bench it gains 0.6 points at 12\% higher cost. As a separate budget-sensitivity analysis, we increase the \textit{Always} budget to $B=10$ (Table~\ref{tab:always-budget}). On SWE-bench, more advice raises success by 20.7 points and brings ER close to \app{}'s, but at higher cost and nearly twice its consultation volume; Terminal-Bench gains only 4.0 points with unchanged ER, while on $\tau^2$-bench success falls and ER turns negative.
\looseness=-1

\textbf{Native orchestrator routing is sparse and late.} \textit{LLM-Router} removes \app's learned policy and lets the native orchestrator agent select the advisor through speaker selection. Despite repeated opportunities to consult, each informed by the full task ledger, the orchestrator requests an average of only 0.41, 0.61, and 0.48 consultations per episode on SWE-bench, Terminal-Bench, and $\tau^2$-bench. Its requests often occur in similar workflow phases as \app but substantially later: 77\%, 64\%, and 76\% of mapped \textit{LLM-Router} consultations occur in the final third, compared with 47\%, 25\%, and 7\% for \app. Native routing therefore identifies similar phases but not \app's timing or consultation frequency.\looseness=-1

\subsection{Consultation price controls the operating point}
\label{sec:lambda}

\begin{figure}[t]
    \vspace{-0.5em}
    \centering
    \includegraphics[width=0.9\textwidth]{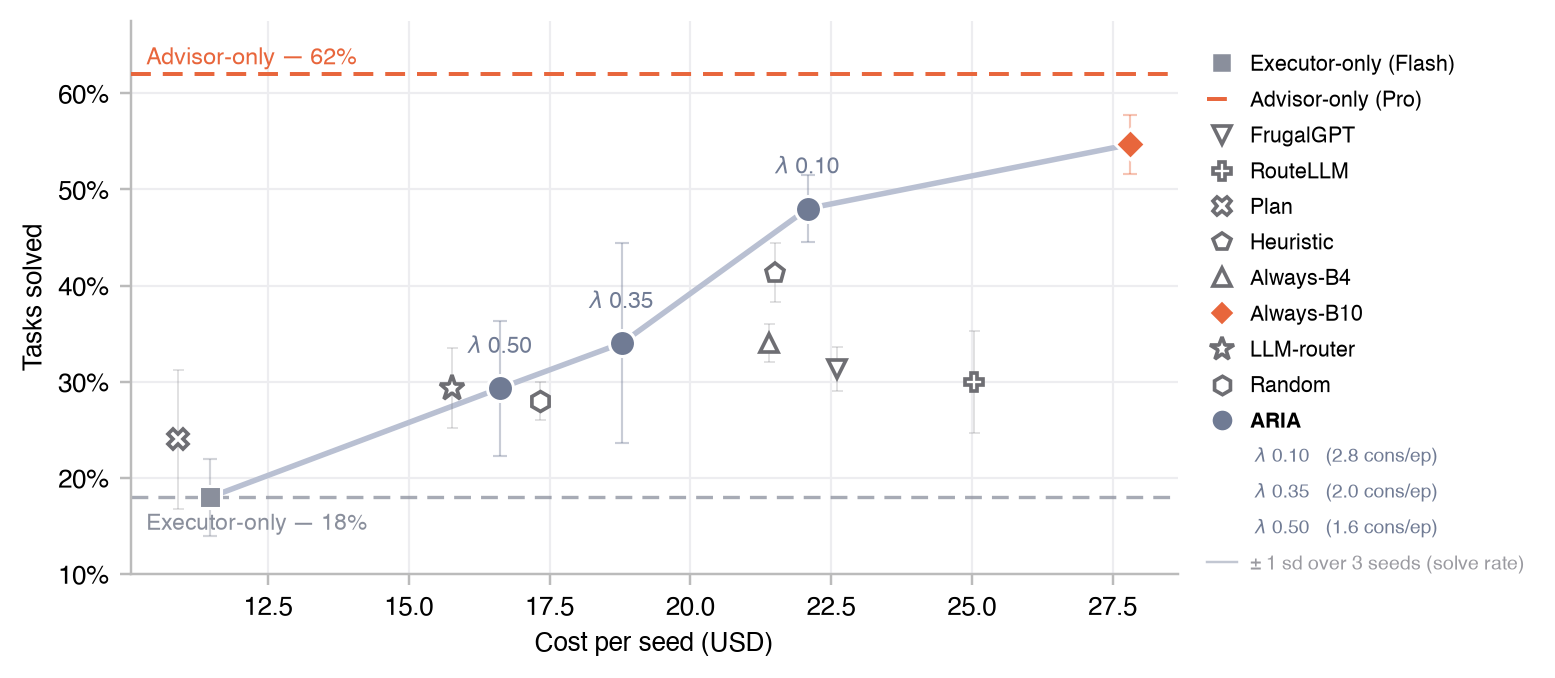}
    \vspace{-0.5em}
    \caption{Success-cost operating points on SWE-bench for \app policies with $\lambda\in\{0.1,0.35,0.5\}$, consultation-budget ablations, and reference conditions.}
    \label{fig:lambda_operating}
    \vspace{-1em}
\end{figure}

\textbf{The consultation price sets \app's operating point.} Table~\ref{tab:main} reports the $\lambda=0.10$ policy, which attains the highest excess recovery among the evaluated conditions on SWE-bench (ER $=0.37$). Figure~\ref{fig:lambda_operating} shows how this point changes as consultation price increases. Increasing $\lambda$ from $0.10$ to $0.35$ and $0.50$ reduces consultation from 2.78 to 2.03 and 1.55 calls per episode and evaluation costs from \$22.09 to \$18.78 and \$16.63, while success falls from 48.0\% to 34.0\% and 29.3\%.

The normalized success-cost view shows that the cheaper policies are not more efficient. ER falls from 0.37 at $\lambda=0.10$ to 0.15 and 0.11 at the higher prices. Thus, $\lambda=0.10$ is the highest-ER operating point among the tested values: the additional consultations recover more performance than the corresponding increase in normalized cost. \textit{Always-$B10$} reaches nearly the same efficiency on SWE-bench (ER $=0.36$) only at roughly twice \app's consultation volume (Section~\ref{sec:placement}). The three \app policies are re-estimated from the same training traces (Section~4), so this price comparison characterizes offline sensitivity to consultation price under the observed visitation distribution rather than independent retraining at each price. As the price increases, \app consults less often and changes where consultation remains worthwhile. Section~\ref{sec:results_policy} examines this structure directly.\looseness=-1

\subsection{Learned allocation differs across workflow domains}
\label{sec:results_policy}
The consultation-price sweep shows how much advice \app{} retains, while
the frozen policy tables reveal where it allocates that advice
(Appendix~\ref{app:policy-diagnostics}). A clear pattern emerges: consultation is concentrated in a small subset of execution states. Each learned policy selects \textsc{Consult} in only 4--9 of the 24 nominal states, so most contact points are left silent by decision rather than by lack of experience. On $\tau^2$-bench, the states not encountered during training are exclusively error-evidence states. Unlike SWE-bench and Terminal-Bench, where coding and terminal actions can expose explicit failures, $\tau^2$ primarily involves interactive dialogue and tool use, so mistakes often remain semantic or procedural rather than producing hard error signals. Its comparatively high baseline success further means that such explicit failure states are rarely encountered. Consultation is thus selective and cannot be explained solely as a response to explicit errors.

Beyond this shared pattern, allocation differs by benchmark: 53.5\% of realized advisor calls on SWE-bench occur at the last temporal stage, \textsc{Deliver}, whereas 82.2\% on $\tau^2$-bench occur during \textsc{Reason}. By contrast, the Anthropic-inspired \textit{Heuristic} condition triggers consultation before committing to an approach and before declaring completion, with additional calls when progress stalls or the approach changes \citep{anthropic_2026_advisor}. These generic triggers do not consistently align with \app's learned consultation patterns. On SWE-bench, our approach places substantially more of its realized consultation at the delivery stage; on $\tau^2$-bench, consultation instead concentrates in the middle of the trajectory. Correspondingly, the \textit{Heuristic} trails \app by 6.7 pp on SWE-bench and 10.0 points on $\tau^2$-bench, while tying it on the more heterogeneous Terminal-Bench. 
Together, these findings suggest that generic consultation rules provide a useful starting point, but benchmark-specific allocation can offer further improvement. Cross-seed policy agreement and a complementary within-state analysis of \textsc{Consult} versus \textsc{Silent} are reported in Appendices~\ref{app:policy_consensus} and~\ref{app:ipw}.

\subsection{Failure analysis reveals mechanism and scope}
\label{sec:results_failures}
To understand how consultation changes the failures that remain, we classify failed trajectories with MAST \citep{cemri_2025_multiagent} and focus here on their primary failure mode (Figure~\ref{fig:mast}); secondary failure modes are reported in Appendix~\ref{app:mast}. On SWE-bench, autonomous execution is dominated by task-verification failures (\textit{Executor-only}: 86\%). As consultation placement becomes more targeted, verification-primary failures account for a smaller share of residual failures: 68\% under \textit{Random}, 46\% under \textit{Heuristic}, and 34\% under \app. \textit{LLM-Router} remains close to the autonomous profile despite access to the same advisor (83\%). In absolute terms, 26 verification-primary failures remain under \app, compared with 70 under \textit{Random}. Consultation therefore changes not only how often the system fails, but also what remains unresolved.

As verification decreases, a different bottleneck emerges. Ignored agent input (FM-2.5) becomes \app's dominant residual mode, accounting for 58\% of its SWE-bench failures. The mode is absent under the orchestrator agent routing (\textit{LLM-Router}: 0\%), less common under uninformed placement (\textit{Random}: 16\%), and increasingly prominent under externally allocated advice (\textit{Heuristic}: 47\%; \textit{Always-$B10$}: 71\%). Across our experiments, FM-2.5 is substantially more prevalent than in MAST-Data (17.3\% vs.\ 1.9\%) \citep{cemri_2025_multiagent}. Human validation confirms that these cases contain relevant advice that reached the executor but was not acted upon (Appendix~\ref{app:mast}). Placement and uptake are therefore distinct problems: \app can learn where advice has high downstream value without ensuring that the executor follows every recommendation. This is evident on SWE-bench at the \textsc{Deliver} stage, where \app often recommends consultation despite the highest ignored-input rate across stages (36--40\%). On $\tau^2$-bench, the ignored-input rate remains below 7\% throughout.\looseness=-1

The same distinction helps explain the cross-benchmark results. 
On Terminal-Bench, \app and \textit{Heuristic} achieve the same success rate, but each solves 12 episodes the other fails. Their remaining failures also differ: verification-primary failures account for 53\% under \app, compared with 82\% under \textit{Heuristic}. Thus, different consultation schedules can produce the same headline performance while succeeding on different tasks and exposing different residual limitations. On $\tau^2$-bench, in contrast, \app leaves only eight judged failures, so the SWE-style uptake bottleneck does not emerge. More generally, MAST profiles depend on the interaction regime as well as on the participating models: communication can remove some failure modes while making others possible. Failure taxonomies should therefore be interpreted as properties of the complete multi-agent system, not of its component models in isolation.

\begin{figure}[H]
    \centering
    \includegraphics[
        width=0.96\linewidth,
        trim=0 5 0 0,
        clip
    ]{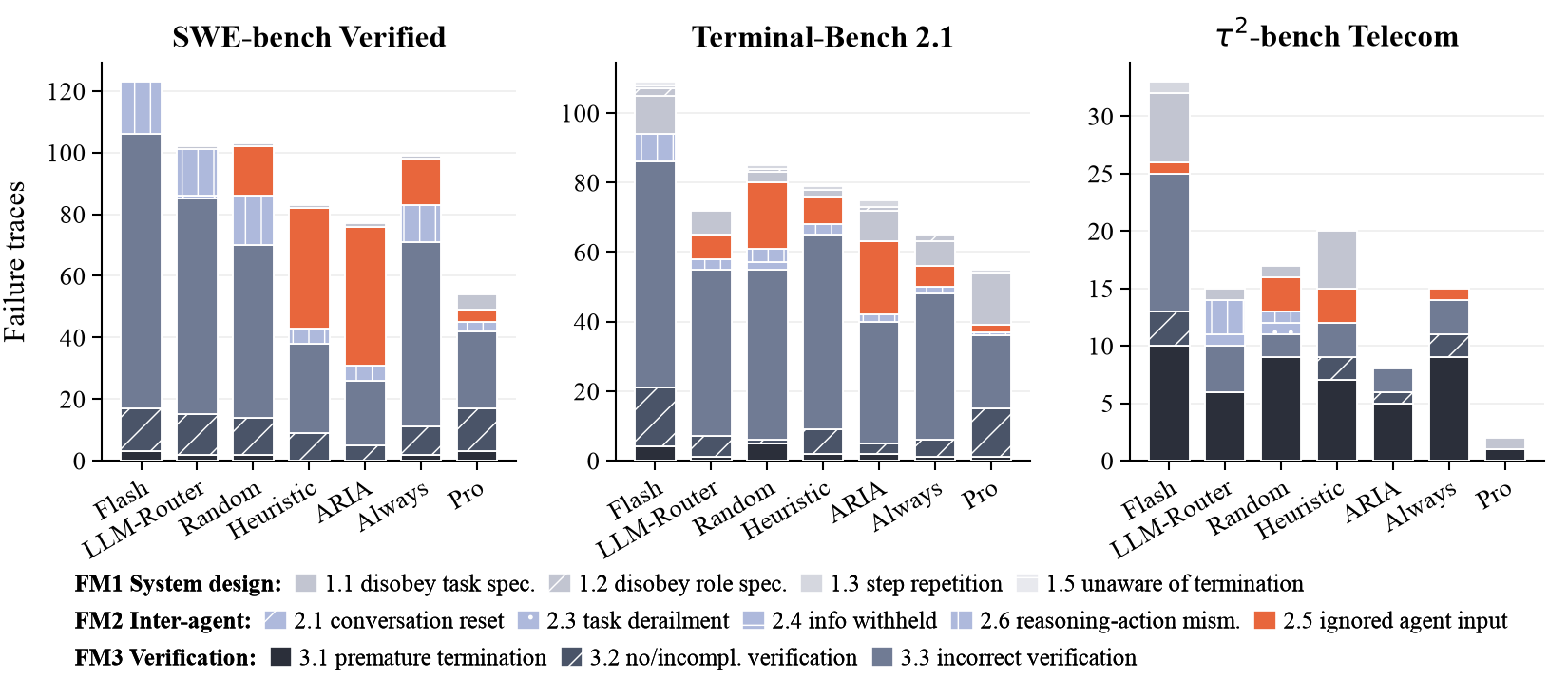}
    \caption{Primary failure mode across baselines and benchmarks grouped into system-design, inter-agent misalignment, and verification failures, showing different residual failure profiles.}
    \label{fig:mast}
\end{figure}


\section{Discussion and Limitations}
Our results show that consultation timing matters more than access or volume alone. \app{} achieves the highest excess recovery on all three benchmarks, outperforming both fixed allocation and self-assessment. Gains are largest when similar decision points repeat across tasks and smaller on heterogeneous tasks such as on Terminal-Bench, suggesting that consultation policies are most useful when learned within related workflows rather than across widely different task types. Although our experiments focus on a single model pair and the Magentic-One framework, ARIA provides a testbed for studying other model pairs, agent configurations, and orchestration frameworks.

Beyond these empirical findings, \app{} makes the learned division of labor inspectable rather than leaving consultation decisions implicit in the underlying models. Its compact policy requires neither fine-tuning nor access to model internals and separates two decisions often conflated in multi-agent systems: when to consult an advisor and what happens once that guidance is provided. Our failure analysis shows that improving the first can expose advice uptake as the next bottleneck, making \app{} both a resource-allocation mechanism and a controlled way to study inter-agent coordination.

\section{Conclusion}
\app{} learns when an executor should consult an advisor during agentic workflows. Across three benchmarks, it recovers 44--81\% of the executor-advisor performance gap at less than half the \textit{Advisor-only} cost and achieves the highest empirical excess recovery among all evaluated conditions. It requires neither fine-tuning nor access to model internals, and represents its decisions as an inspectable policy with a single parameter controlling the consultation trade-off. Together, these results establish selective consultation as an effective mechanism for executor-advisor collaboration, while the failure analysis identifies advice uptake as a further lever for improving efficiency.

\subsection*{AI use statement}

Generative AI tools, including ChatGPT (Astra) and Claude (Fable 5, Opus 5), were used during the preparation of this work to support revision, literature discovery and organization, research ideation, methodological cross-checking, technical implementation and analysis support, figure preparation, and the drafting of selected manuscript sections. AI-assisted outputs were treated as suggestions rather than authoritative sources and were reviewed, edited, and verified by the authors before inclusion. All citations, quantitative results, experimental claims, benchmark splits, tables, figures, and methodological descriptions were checked against the underlying literature, implementation, experiment logs, or source artifacts. The authors take full responsibility for the final content of the paper.

\subsection*{Reproducibility statement}
The consultation objective, state construction, Monte Carlo updates, exploration schedule, training budgets, and frozen deployment policy are specified in Section~3 and Appendix~A. Appendix~B documents the benchmark splits, baseline implementations, \textit{Random}-control calibration, model settings, and token- and cost-accounting procedure. Appendix~C reports seed-level variation, controller diagnostics, policy agreement, within-state analyses, and the MAST failure-analysis protocol. All headline results are averaged over three evaluation seeds, with corresponding dispersion reported separately. The code repository provides the implementation.\footnote{\url{https://github.com/Pasl4/aria}}

\bibliography{main}
\bibliographystyle{iclr2026_conference}

\clearpage
\appendix
\section{Methodological Details}
\label{app:method-details}

\subsection{Advisor Intervention Semantics} 
\label{app:advisor-semantics}

Section~\ref{sec:placing_decision} defines consultation as a binary decision at recurring workflow contact points. Here, we specify the intervention contract used once consultation is triggered. Across \app, \textit{Random}, \textit{Heuristic}, \textit{Always}, and \textit{Plan}, this contract is held fixed; the conditions differ only in the rule that determines which eligible contact points trigger consultation.

When a condition selects \textsc{Consult}, the advisor has access to the task specifications and the full available trajectory context without additional summarization or truncation, and returns natural-language guidance with recommended next steps. The advisor is read-only: it does not execute tools, take environment actions, or submit the final answer. The executor remains responsible for all subsequent reasoning, tool use, environment interaction, and final submission, and may follow, adapt, or ignore the recommendation. Consultation therefore provides additional reasoning without transferring execution control to the advisor.

\paragraph{Contact-point-specific intervention.}
How the guidance enters the trajectory is determined solely by the contact-point type defined in Section~\ref{sec:placing_decision}. At \texttt{PRE\_STEP}, the advisor response is appended to the orchestrator's next instruction before execution continues. At \texttt{STALL}, guidance is inserted before replanning so that the revised plan can incorporate the advisor's assessment. At \texttt{PRE\_DELIVERY}, the advisor reviews the current solution before final submission; if it flags the trajectory as off track, the workflow permits one additional revision cycle before finalization. These insertion rules are identical across the consultation conditions above.

\paragraph{Silent action and downstream effects.}
Under \textsc{Silent}, no advisor call is made and the underlying workflow continues unchanged. Thus, the meta-controller's direct intervention is limited to whether advisor guidance enters the trajectory; the executor, advisor, task interface, and surrounding workflow otherwise remain fixed. Once introduced, however, advice can indirectly change later executor actions, observations, and contact points through the executor's response to that guidance. The \textit{LLM-Router} baseline follows a separate routing contract, described in Appendix~\ref{app:baselines}. This separation allows consultation allocation and advice uptake to be analyzed independently.

\subsection{State Construction}
\label{app:state_construction}
Section~\ref{sec:states} defines the compact state abstraction $s_t=(p_t,e_t,h_t)$. Here, we specify the exact mapping from workflow events and observable execution feedback to each state component.

\paragraph{Design principle.}
\app maps the observable workflow history $H_t$ to the compact state $s_t=(p_t,e_t,h_t)$ defined in Section~\ref{sec:states}. The representation is intentionally based on workflow-level signals rather than model-internal confidence or self-assessment, whose reliability is known to be limited \citep{tyen_2024_errors,xiong_2024_uncertainty}. The abstraction is designed to pool consultation experience across recurring execution contexts rather than reconstruct the full trajectory. This follows the general motivation of approximate state abstraction: a smaller representation improves reuse and interpretability at the cost of discarding potentially decision-relevant information \citep{abel_2016_near}.

\paragraph{Temporal-state mapping.}
Magentic-One provides recurring orchestration events during execution \citep{fourney_2024_magenticone}. We use three such events as consultation
contact points: \texttt{PRE\_STEP}, before the next executor instruction;
\texttt{STALL}, when stalled progress triggers replanning; and
\texttt{PRE\_DELIVERY}, immediately before final delivery. Contact-point type
and temporal state are distinct: the former identifies why a consultation
opportunity occurs, whereas $p_t$ provides a coarse representation of where
that opportunity lies in the trajectory.

In the experimental setup used throughout this paper, the temporal state is assigned as
\[
p_t =
\begin{cases}
\textsc{Frame},   & \text{if } \mathrm{kind}_t \neq \texttt{PRE\_DELIVERY}
                      \text{ and } i_t = 0,\\
\textsc{Explore}, & \text{if } \mathrm{kind}_t \neq \texttt{PRE\_DELIVERY}
                      \text{ and } i_t \in \{1,2\},\\
\textsc{Reason},  & \text{if } \mathrm{kind}_t \neq \texttt{PRE\_DELIVERY}
                      \text{ and } i_t \geq 3,\\
\textsc{Deliver}, & \text{if } \mathrm{kind}_t = \texttt{PRE\_DELIVERY},
\end{cases}
\]
where $i_t$ is the zero-based index over all contact points in the episode. All reported experiments use this fixed mapping. Although the concrete mapping is specific to Magentic-One, the abstraction is designed to generalize: other single- or multi-agent systems can instantiate the same idea using their own recurring workflow events and corresponding progress signals. The controller records the index before constructing the state and increments it afterwards. Consequently, \texttt{STALL} creates an additional decision opportunity but does not define a temporal stage of its own. Its temporal label is determined by the same ordinal rule as other non-delivery contact points. The labels are inspired by coarse stages of problem solving \citep{polya_1945_solve}; they should be interpreted as task-agnostic progress labels rather than semantic classifications of the executor's reasoning.

\paragraph{Execution-evidence mapping.}
The evidence dimension summarizes observable execution feedback. Such feedback can support correction and self-repair \citep{chen_2024_selfdebug, shinn_2023_reflexion}, although its usefulness depends on which signals are exposed by the environment \citep{olausson_2024_repair, huang_2024_selfcorrect}. Accordingly, $e_t$ records execution status rather than semantic correctness.

For SWE-bench and Terminal-Bench, the implementation considers the most recent
step containing executed output. If no such step exists,
$e_t=\textsc{None}$. Otherwise, the step maps to \textsc{Error} when its
execution-error field is set and to \textsc{Clean} otherwise. The error field
is derived from observable execution output using explicit failure signals, including tracebacks or exceptions, non-zero exit codes, explicit failure markers, and non-zero pytest failure summaries. The most recent step determines the state, so a failure followed by a successful repair does not remain encoded as \textsc{Error}. 

For $\tau^2$-bench, the interaction environment exposes tool results rather than conventional process output. A step is therefore treated as acting when it produces either executed output or a harness-recorded tool result. If no acting step exists, $e_t=\textsc{None}$; explicit machine- or tool-reported
errors map to the failure state, and other tool results map to the clean state. The implementation names these two non-empty values \textsc{Ok} and \textsc{Failed}; throughout the paper they correspond to \textsc{Clean} and \textsc{Error}, respectively. End-of-conversation notifications are not encoded as execution failures.

\paragraph{Consultation-history mapping.}
The final state component, $h_t\in\{0,1\}$, records whether advisor guidance has already entered the trajectory. This distinction allows the controller to represent that the value of a further consultation may differ once the executor has already received advice. Remaining consultation budget is deliberately omitted from the state; once the episode-level budget $B$ is exhausted, subsequent \textsc{Consult} choices are forced to \textsc{Silent}. Together, these mappings produce the 24 state representation used by the controller. The abstraction preserves three factors that can systematically change consultation value: trajectory progress, observable execution evidence, and prior advisor access, while remaining small enough for direct tabular estimation and inspection.

Together, these deterministic mappings make every deployment state directly recoverable from observable workflow signals while keeping the meta-controller small enough for tabular estimation and inspection.

\subsection{Training and Policy Estimation}
\label{app:training}

\app{} learns only the consultation controller; the executor and advisor
remain fixed throughout training. This section details how the cost-aware
objective in Eq.~\ref{eq:aria-return} is used to estimate action values and
derive the frozen deployment policy.

\paragraph{Training objective and cost calibration.}
Each consultation incurs the fixed penalty
$\kappa=\lambda\bar{c}/\Delta$, where $\bar{c}$ is the mean consultation cost
and $\Delta$ is the mean per-episode spend of consulting at every contact
point. All headline policies use $\lambda=0.10$.
Table~\ref{tab:cost-calibration} reports the benchmark-specific calibration
values and the resulting penalties. These penalties define the learning
objective; realized evaluation costs are reported separately.

\begin{table}[H]
    \centering
    \small
    \caption{Benchmark-specific cost calibration for the headline policies
    at $\lambda=0.10$. The quantities $\bar{c}$ and $\Delta$ are in USD;
    $\kappa=\lambda\bar{c}/\Delta$ is the per-consultation penalty in
    reward units.}
    \label{tab:cost-calibration}
    \begin{tabular}{lrrr}
        \toprule
        Benchmark & $\bar{c}$ & $\Delta$ & $\kappa$ \\
        \midrule
        SWE-bench
            & 0.0490 & 0.1663 & 0.029465 \\
        Terminal-Bench
            & 0.0683 & 0.2816 & 0.024254 \\
        $\tau^2$-bench
            & 0.0244 & 0.1281 & 0.019048 \\
        \bottomrule
    \end{tabular}
\end{table}

\paragraph{Action-value estimation.}
We use every-visit tabular Monte Carlo control with sample-average updates
and $\gamma=1$. At contact point $t$ in trajectory $\tau$, the update target
is the return-to-go
\[
    G_t = Y(\tau) - \kappa N_c^{\geq t}(\tau),
\]
where $Y(\tau)$ is terminal task success and $N_c^{\geq t}(\tau)$ counts
consultations at or after that visit. Only episodes reaching a terminal
task outcome contribute value updates. For each visited state-action pair
$(s,a)$,
\begin{equation}
    Q(s,a)
    =
    \frac{1}{N(s,a)}
    \sum_{j=1}^{N(s,a)} G_j ,
    \label{eq:mc-sample-average}
\end{equation}
where $N(s,a)$ is the number of visits to $(s,a)$ and $G_j$ is the
return-to-go from the corresponding visit.

Every occurrence contributes one sample, retaining all observed
state-action returns in the compact state space. Repeated visits within
an episode therefore contribute correlated samples, and long trajectories
can receive more weight than under first-visit Monte Carlo.
The equivalent incremental learning rate is $\alpha=1/N(s,a)$, so no
additional learning-rate hyperparameter is introduced.

\paragraph{Exploration.}
Training follows an $\epsilon$-greedy behavior policy over
$\{\textsc{Silent},\textsc{Consult}\}$. Let $m$ index training episodes,
distinct from the contact-point index $t$. Exploration is annealed
linearly from $\epsilon_0=0.40$ to $\epsilon_{\min}=0.10$ over the first
110 training episodes:
\begin{equation}
    \epsilon(m)
    =
    \max\!\left(
        0.10,\;
        0.40 - 0.30\frac{m}{110}
    \right).
    \label{eq:epsilon-schedule}
\end{equation}
The exploration rate remains at $0.10$ thereafter. This schedule is
identical across benchmarks and seeds and was verified against the
stored checkpoints of all nine training runs.

\paragraph{Training budgets.}
We train three independent controller seeds per benchmark.
Table~\ref{tab:training-config} summarizes the training budgets and
shared hyperparameters. Development tasks are revisited to accumulate
controller experience, corresponding to approximately three, six, and
eight passes on SWE-bench, Terminal-Bench, and $\tau^2$-bench,
respectively. Training and held-out evaluation tasks are disjoint;
Appendix~\ref{app:splits} specifies the partitions.

\begin{table}[htbp]
    \centering
    \small
    \caption{Controller-training budgets and shared hyperparameters.
    Evaluation uses a frozen greedy policy without exploration.}
    \label{tab:training-config}
    \begin{tabular}{lccc}
        \toprule
        & \textbf{SWE-bench}
        & \textbf{Terminal-Bench}
        & $\boldsymbol{\tau^2}$\textbf{-bench} \\
        \midrule
        Controller seeds
            & 3 & 3 & 3 \\
        Training episodes
            & $\sim$150 & 162 & 160 \\
        Training tasks
            & $\sim$50 & 27 & 20 \\
        Approx.\ passes
            & 3 & 6 & 8 \\
        Discount $\gamma$
            & 1 & 1 & 1 \\
        MC update
            & \multicolumn{3}{c}{$\alpha=1/N(s,a)$} \\
        $\epsilon_0$
            & \multicolumn{3}{c}{0.40} \\
        $\epsilon_{\min}$
            & \multicolumn{3}{c}{0.10} \\
        Annealing
            & \multicolumn{3}{c}{linear, first 110 episodes} \\
        Evaluation $\epsilon$
            & \multicolumn{3}{c}{0 (greedy)} \\
        \bottomrule
    \end{tabular}
\end{table}

\paragraph{Offline consultation-price re-estimation.}
The SWE-bench consultation-price ablation reuses the training trajectories
collected under the $\lambda=0.10$ behavior policy.
For $\lambda\in\{0.35,0.50\}$, we recompute the visit-specific returns
with the corresponding consultation penalty, rebuild the sample-average
value table, and evaluate the resulting greedy policy on fresh
trajectories. This changes the consultation price without modifying
either language model or collecting additional training episodes.
The resulting tables are offline re-estimates under the original
behavior-policy visitation distribution, not independently trained
policies at each price.

\paragraph{Frozen deployment policy.}
After training or offline re-estimation, the value table is frozen.
Evaluation selects the greedy action in each state with $\epsilon=0$;
ties, including entirely unvisited rows, resolve to \textsc{Silent}.
The episode-level consultation budget is enforced separately: once it
is exhausted, subsequent \textsc{Consult} choices are forced to
\textsc{Silent}, as described in
Appendix~\ref{app:state_construction}.
Deployment therefore requires only deterministic state extraction,
table lookup, and budget enforcement, with no exploration or controller
updates on held-out tasks.

\clearpage
\section{Experimental Details}
\label{app:experimental-details}

\subsection{Split Construction and Task Separation}
\label{app:splits}

For each benchmark, controller development and final evaluation use
disjoint task instances from the same benchmark family. The executor and
advisor remain fixed throughout; only \app's consultation policy is learned
on the development tasks. Final evaluation uses the same held-out task set
across all compared methods.

Table~\ref{tab:data-splits} summarizes the resulting development and
evaluation partitions.

\begin{table}[H]
    \centering
    \footnotesize
    \setlength{\tabcolsep}{4pt}
    \caption{
        Development and evaluation splits used for policy learning and final
        evaluation. Train--test overlap was verified from task identifiers
        rather than assumed from split names.
    }
    \label{tab:data-splits}

    \begin{tabular*}{\textwidth}{
        @{\extracolsep{\fill}}
        l l c c l
        @{}
    }
        \toprule
        \textbf{Benchmark}
        & \textbf{Development source}
        & \textbf{Train tasks}
        & \textbf{Train episodes}
        & \textbf{Held-out test} \\
        \midrule

        SWE-bench
            & screened benchmark training pool
            & 50-52
            & 150-156
            & 50 (Verified-Mini) \\

        Terminal-Bench
            & 39-task development partition
            & 27
            & 162
            & 50 (test) \\

        $\tau^2$-bench
            & official telecom train split
            & 20
            & 160
            & 40 (telecom test) \\

        \bottomrule
    \end{tabular*}
\end{table}

\paragraph{SWE-bench Verified.}
The development pool was constructed before controller training from a
screening of 100 candidate tasks drawn from the SWE-bench Verified benchmark outside the Mini evaluation subset. Both underlying models were first run autonomously on these candidates. We retained two screening strata: (A) tasks solved by the \textit{Executor-only} and (B) tasks solved by the \textit{Advisor-only}. This initially produced 52 development tasks, comprising 21 tasks from stratum A and 31 from stratum B. The first controller seed used all 52 tasks; following unrecoverable rerun crashes, subsequent seeds were standardized to a 50-task development set. Tasks solved by neither model were excluded from controller development. The resulting development set is therefore not outcome-blind: its construction intentionally retains tasks on which autonomous execution is feasible or stronger-model capability provides a demonstrated advantage. The held-out 50-task Verified-Mini evaluation set was not used in this screening procedure, and its intersection with the 52-task development pool is empty.

\paragraph{Terminal-Bench 2.1.}
We first partitioned all 89 tasks into a 39-task development partition and a 50-task held-out test partition. The development partition was then filtered using six completed autonomous benchmark runs, comprising three runs with the executor model and three with the advisor model. Twelve development tasks that were not solved in any of these six runs were removed, leaving 27 controller development tasks. The held-out 50-task test partition was left unchanged. Each of the 27 development tasks is revisited for six passes, yielding 162 training episodes per controller seed.

This filtering is also not outcome-blind. In particular, it removes tasks that were never solved by either model from the development set while retaining such tasks in the held-out evaluation set. Consequently, the development set contains no never-solved tasks, whereas 15 of the 50 held-out tasks (30\%) fall into this category. This makes the development distribution easier than
the evaluation distribution and can increase consultation pressure when the frozen policy is deployed on the harder held-out tasks. The resulting asymmetry therefore does not favor lower evaluation cost.

\paragraph{$\tau^2$-bench Telecom.}
We use 20 tasks from the official Telecom training split for controller training and the official 40-task Telecom test split verbatim for final evaluation. No evaluation task is used during controller learning.

\paragraph{Task separation and repeated development episodes.}
The development tasks are revisited to accumulate controller experience: approximately three passes on SWE-bench, six passes on Terminal-Bench, and eight passes on $\tau^2$-bench. These repetitions provide additional state-action observations but do not introduce held-out test instances into policy learning.

We explicitly verified task identities after data preparation. The measured training-test overlap is zero for all three benchmarks: 0/50 held-out tasks for SWE-bench, 0/50 for Terminal-Bench, and 0/40 for $\tau^2$-bench. Thus, all reported final-evaluation results are obtained on task instances that were not used to derive the frozen consultation tables.

Training episodes that terminate before reaching a terminal task outcome do not update the controller, consistent with Appendix~\ref{app:training}. During final evaluation, failed or crashed episodes remain part of the benchmark denominator even when no usable trajectory is available for subsequent diagnostic analyses.

\subsection{Baseline Contracts}
\label{app:baselines}

All evaluation conditions use the same held-out task instances, task order,
executor--advisor model pair, workflow implementation, and benchmark harness
unless the allocation method itself requires different execution semantics.
The lower-cost executor is \texttt{gemini-3-flash-preview} at low reasoning
effort, and the stronger model is \texttt{gemini-3.1-pro-preview} at high
reasoning effort. Within-trajectory consultation conditions use the common
advisor interface defined in Appendix~\ref{app:advisor-semantics} unless
stated otherwise.

Headline consultation conditions use an episode-level budget of $B=4$
advisor calls. \textit{Plan} consults only at the first eligible opportunity
and therefore has an effective budget of $B=1$. Capability anchors and
task-level allocation baselines do not use this within-trajectory
consultation budget. Table~\ref{tab:baseline-contracts} summarizes the
resulting contracts.

\begin{table}[H]
    \centering
    \small
    \setlength{\tabcolsep}{7pt}
    \renewcommand{\arraystretch}{1.05}

    \caption{
        Baseline contracts. $B$ denotes the maximum number of
        within-trajectory advisor consultations per episode and therefore
        applies only to consultation-based conditions.
    }
    \label{tab:baseline-contracts}

    \begin{tabular}{@{}llc@{}}
        \toprule
        \textbf{Method}
        & \textbf{Allocation rule}
        & \textbf{Budget} \\
        \midrule

        Executor-only
        & no advisor
        & -- \\

        Advisor-only
        & advisor executes
        & -- \\

        FrugalGPT
        & score-based cascade
        & -- \\

        RouteLLM
        & pre-task routing
        & -- \\

        AutoMix
        & verify then escalate
        & -- \\

        Plan
        & first contact
        & $B=1$ \\

        Heuristic
        & beginning/end triggers
        & $B=4$ \\

        Always
        & at every contact point
        & $B=4$ \\

        LLM-Router
        & orchestrator agent selects advisor
        & $B=4$ \\

        Random
        & rate-matched advice
        & $B=4$ \\

        \app{}
        & state-conditioned policy
        & $B=4$ \\

        \bottomrule
    \end{tabular}
\end{table}

\paragraph{Capability anchors.}
\textit{Executor-only} and \textit{Advisor-only} define the two autonomous
reference points used throughout the evaluation. Under \textit{Executor-only},
the lower-cost executor completes the workflow without access to advisor
guidance. Under \textit{Advisor-only}, the stronger model replaces the
executor and completes the full workflow itself. These conditions therefore
vary model capability rather than consultation placement and define the
success and cost anchors used for PGR, CostShare, and ER.

\paragraph{Task-level allocation baselines.}
\textit{FrugalGPT}, \textit{RouteLLM}, and \textit{AutoMix} allocate model
capability at task level rather than at recurring workflow contact points.
\textit{FrugalGPT} implements a score-based cascade: every task first incurs
lower-cost executor execution, and tasks selected for escalation additionally
incur stronger-model execution. \textit{RouteLLM} performs routing before
agentic execution and selects exactly one of the two models to execute the
task. \textit{AutoMix} first executes the task with the lower-cost model,
applies its self-verification step, and invokes stronger-model execution for
tasks that trigger escalation. The self-verification pass is metered
separately and included in the reported \textit{AutoMix} cost. These
conditions test whether task-level routing or cascading is sufficient without
conditioning allocation on evidence revealed later in the trajectory.

\paragraph{Fixed consultation schedules.}
\textit{Plan}, \textit{Heuristic}, and \textit{Always} use the same
executor--advisor intervention contract as \app{} but replace its learned
allocation rule with fixed schedules. \textit{Plan} consults only at the first
eligible contact point and therefore tests front-loaded advisor access.
\textit{Heuristic} uses fixed beginning/end and stall-related triggers.
\textit{Always} requests advice at every eligible contact point until the
episode-level budget is exhausted, providing the high-consultation reference
under the shared $B=4$ cap.

\paragraph{Native orchestrator routing.}
\textit{LLM-Router} removes \app's learned consultation policy and lets the
native LLM orchestrator decide when to involve the advisor. In the underlying
workflow, the orchestrator already selects which available agent acts next;
for this condition, the advisor is included among those selectable agents.
The orchestrator therefore receives repeated opportunities to involve the
advisor during execution, with each decision informed by the full task ledger
available at that point. It does not use \app's compact state abstraction or
learned value table. Advisor selections count against the same episode-level
budget of $B=4$ used by the other headline consultation conditions.
\textit{LLM-Router} thus tests whether the workflow's native LLM orchestrator
can determine when to involve the advisor without a separately learned
consultation controller.

\paragraph{Placement control.}
\textit{Random} replaces \app's state-conditioned action with an independent
Bernoulli decision at every eligible contact point while retaining the same
advisor intervention semantics and episode-level budget. Its probability is
calibrated before running the control from the corresponding fixed \app{}
reference arm to approximate its consultation volume. The complete
calibration procedure and realized matching rates are reported in
Appendix~\ref{app:random-calibration}.

\paragraph{Learned consultation.}
\app{} uses the frozen greedy consultation policy learned on the development
tasks. At each eligible contact point, it maps the observable workflow state
to \textsc{Silent} or \textsc{Consult}; ties, including unvisited states,
resolve to \textsc{Silent}. The advisor interface itself is identical to that
used by \textit{Plan}, \textit{Heuristic}, \textit{Always}, and
\textit{Random}. These conditions therefore differ in the rule that determines
where consultation is allocated rather than in the form of advice delivered.

\paragraph{Budget sensitivity.}
For the separate budget-sensitivity analysis in Section~5.1, we additionally
evaluate \textit{Always} with $B=10$ on all three benchmarks. These variants
are reported separately and are not part of the common-budget headline
comparison.

\subsection{Random Calibration}
\label{app:random-calibration}

\textit{Random} serves as a placement control for \app{}. At each eligible
contact point, it consults the advisor independently with a fixed probability
$p$, subject to the same episode-level budget of $B=4$. Unlike \app{}, the
decision does not depend on the current state or trajectory. The probability
$p$ is fixed before evaluating \textit{Random} and is chosen to approximately
match the consultation volume of the corresponding headline \app{} policy
($\lambda=0.10$).

Let $K_i$ denote the number of consultations made by the \app{} reference
policy in episode $i$, and let $C_i$ denote the number of eligible contact
points in that episode. The nominal consultation probability is

\begin{equation}
    p =
    \frac{\sum_i K_i}
         {\sum_i C_i}.
    \label{eq:random-calibration-rate}
\end{equation}

The resulting value is rounded to two decimal places, or three when
$p<0.25$, and is then held fixed throughout evaluation.

For Terminal-Bench, the $B=4$ consultation cap can noticeably reduce the
number of consultations realized from independent draws. In addition,
complete contact-point logs are unavailable for 25 of the 150 crash-path
episodes. We therefore use reconstructed contact-point counts and choose $p$
so that the expected capped consultation volume matches the \app{} reference:

\begin{equation}
    \sum_i
    \mathbb{E}\!\left[
        \min\!\left(\operatorname{Bin}(C_i,p),B\right)
    \right]
    =
    \sum_i K_i.
    \label{eq:random-calibration-capped}
\end{equation}

Table~\ref{tab:random-calibration} reports the resulting probabilities and
realized consultation volumes.

\begin{table}[H]
    \centering
    \small
    \caption{
        Calibration and realized consultation volume of the Random placement
        control relative to the headline \app{} policy ($\lambda=0.10$).
    }
    \label{tab:random-calibration}

    \begin{tabular}{lccc}
        \toprule
        & \textbf{SWE-bench}
        & \textbf{Terminal-Bench}
        & $\boldsymbol{\tau^2}$\textbf{-bench} \\
        \midrule

        \app consults/episode
            & 2.780
            & 2.460
            & 3.042 \\

        Consultation probability $p$
            & 0.49
            & 0.62
            & 0.172 \\

        Random consults/episode
            & 2.527
            & 2.417
            & 2.670 \\

        Random / \app volume
            & 90.9\%
            & 98.2\%
            & 87.8\% \\

        \bottomrule
    \end{tabular}
\end{table}

We consider consultation volume matched when the realized number of
consultations per episode is within $\pm10\%$ of the \app{} reference.
SWE-bench and Terminal-Bench satisfy this criterion. On $\tau^2$-bench,
\textit{Random} reaches 87.8\% of the \app{} consultation volume and is
therefore treated as approximately rate-matched rather than exactly matched.

We match consultation volume rather than dollar cost because consultation
timing can change the subsequent trajectory and therefore the amount of context
processed by later model calls. Realized model cost is reported separately.
The probability $p$ is not adjusted after observing the \textit{Random}
evaluation trajectories.

\subsection{Token Usage and Billing}
\label{app:tokens}

To make the reported cost comparisons independently auditable, we reconstruct
the headline evaluation costs from the raw billing artifacts and reconcile them
against both the episode ledgers and the values reported in Table~2. The scope
is restricted to the headline configuration: three evaluation seeds for each
of the 33 benchmark-condition cells, with the common $B=4$ consultation
budget where applicable. Training runs, pilot experiments, the
$\lambda\in\{0.35,0.50\}$ ablations, and the $B=10$ budget-sensitivity arms are
excluded. All 33 headline cells are recovered, with reconstructed mean costs
differing from Table~2 by at most \$0.0049.

\paragraph{Pricing and token accounting.}
Each model call records total prompt tokens, cached prompt tokens, completion
tokens, reasoning tokens, and billed cost. Because the prompt count includes
the cached portion, fresh input is
\[
    f = n_{\mathrm{prompt}} - n_{\mathrm{cached}}.
\]
For a call with fresh input $f$, cached input $k$, completion tokens $o$, and
reasoning tokens $r$, the base cost is
\[
    c =
    \frac{
        f p_{\mathrm{in}}
        + k p_{\mathrm{cached}}
        + (o+r)p_{\mathrm{out}}
    }{10^6}.
\]
Reasoning tokens are therefore billed in addition to completion tokens at the
output-token rate. This interpretation reproduces the logged provider cost for
all 106{,}662 non-failed priced calls in the reconstruction; treating reasoning
tokens as already contained in the completion count does not.

The pricing schedule is read from the frozen experiment configuration and
mirrored into each run manifest. For
\texttt{gemini-3.1-pro-preview}, calls whose prompt exceeds 200K tokens incur
the long-context surcharge shown in Table~\ref{tab:pricing}. The surcharge is
applied per call after the base logger calculation.

\begin{table}[H]
    \centering
    \small
    \resizebox{\textwidth}{!}{%
\begin{tabular}{lrrrrll}
\toprule
Model & Input & Cached input & Output & Reasoning & Tier threshold & Above threshold \\
\midrule
\texttt{gemini-3-flash-preview} & 0.50 & 0.05 & 3.00 & 3.00 & none & none \\
\texttt{gemini-3.1-pro-preview} & 2.00 & 0.20 & 12.00 & 12.00 & 200,000 prompt tokens & in 2.00 / cached 0.20 / out 6.00 \\
\bottomrule
\end{tabular}

    }
    \caption{Pricing used for the reported evaluation costs, in USD per
    million tokens. The final column reports the incremental surcharge applied
    above the listed context threshold. Thus, for
    \texttt{gemini-3.1-pro-preview}, the effective rates above 200K prompt
    tokens are \$4.00 for fresh input, \$0.40 for cached input, and \$18.00 for
    output and reasoning tokens.}
    \label{tab:pricing}
\end{table}

\paragraph{Episode and condition attribution.}
A re-run episode may leave multiple attempts in the billing log while the
episode ledger retains only the attempt used for evaluation. We identify the
attributed attempt by matching the sum of its call costs to the corresponding
ledger cost rather than by file order. Within the headline audit scope,
4{,}066 unique episode records contain at least one call record. Of these,
4{,}052 contain a single matching attempt and 14 contain multiple attempts
with exactly one match; no case remains unresolved.

The accounting also follows the execution semantics of each condition.
\textit{FrugalGPT} and \textit{AutoMix} are cascades and therefore incur the
lower-cost executor run for every task plus stronger-model execution for the
tasks they escalate. \textit{RouteLLM} selects exactly one executor per task.
The \textit{AutoMix} self-verification pass is metered separately from the
agent call log and is included in its reported cost. Repeated task runs are
resolved according to the frozen campaign precedence before aggregation.

\paragraph{Benchmark-side model usage.}
The $\tau^2$ user simulator is metered separately from the evaluated agent.
It is part of the benchmark environment and is not included in the model costs
reported in Table~2. Across the reconstructed $\tau^2$ source runs, this
simulator accounts for 24{,}826 calls, 110.6M recorded tokens, and \$48.26.
Its meter does not expose cached-input tokens, so cached simulator usage is
reported as unavailable rather than inferred or treated as zero.

\paragraph{Token usage.}
Table~\ref{tab:token-summary} reports the complete token breakdown used for
the headline conditions. Fresh input and cached input are shown separately,
and reasoning tokens are reported separately from ordinary completion tokens.
All token quantities are summed across the three evaluation seeds, whereas
the final cost column reports the mean evaluation-run cost used in Table~2.

For cascade conditions, the component-episode count includes both the base
execution and any escalated execution and should therefore not be interpreted
as the number of unique benchmark tasks. Likewise, rows in
Table~\ref{tab:token-summary} must not be summed to obtain a campaign total:
some conditions deliberately reuse and re-charge the same lower-cost base runs.

\begin{table}[H]
    \centering
    \scriptsize
    \resizebox{\textwidth}{!}{%
\begin{tabular}{llrrrrrrrrr}
\toprule
Benchmark & Condition & Ep. & Calls & Input & Cached & Output & Reason. & Total & Cached \% & Cost/run \\
\midrule
SWE-bench Verified & Executor-only & 150 & 2,440 & 36.825 & 12.397 & 4.251 & 0.880 & 54.354 & 25.2 & 11.48 \\
SWE-bench Verified & Advisor-only & 150 & 3,367 & 28.609 & 43.244 & 2.164 & 2.886 & 76.903 & 60.2 & 45.90 \\
SWE-bench Verified & FrugalGPT & 196 & 3,470 & 43.958 & 21.112 & 4.850 & 1.675 & 71.596 & 32.5 & 22.61 \\
SWE-bench Verified & RouteLLM & 150 & 2,734 & 29.913 & 21.392 & 3.524 & 1.924 & 56.753 & 41.7 & 25.03 \\
SWE-bench Verified & AutoMix & 257 & 4,945 & 66.469 & 66.846 & 5.954 & 3.133 & 144.438 & 50.1 & 59.68 \\
SWE-bench Verified & Plan & 150 & 2,736 & 31.402 & 9.571 & 3.782 & 0.435 & 45.190 & 23.4 & 10.90 \\
SWE-bench Verified & Heuristic & 150 & 3,453 & 47.571 & 43.610 & 4.446 & 1.655 & 97.281 & 47.8 & 21.50 \\
SWE-bench Verified & Always & 150 & 3,827 & 44.080 & 38.790 & 3.908 & 2.110 & 88.887 & 46.8 & 21.39 \\
SWE-bench Verified & LLM-Router & 150 & 2,870 & 46.561 & 37.974 & 4.950 & 1.145 & 90.630 & 44.9 & 15.77 \\
SWE-bench Verified & Random & 150 & 3,320 & 38.589 & 23.170 & 4.085 & 1.485 & 67.329 & 37.5 & 17.33 \\
SWE-bench Verified & ARIA & 150 & 3,488 & 46.150 & 39.643 & 4.729 & 1.585 & 92.107 & 46.2 & 22.09 \\
\midrule
Terminal-Bench 2.1 & Executor-only & 150 & 1,827 & 31.581 & 17.717 & 3.512 & 0.347 & 53.157 & 35.9 & 9.76 \\
Terminal-Bench 2.1 & Advisor-only & 150 & 3,332 & 36.416 & 83.526 & 1.860 & 3.658 & 125.460 & 69.6 & 60.99 \\
Terminal-Bench 2.1 & FrugalGPT & 154 & 1,919 & 32.161 & 18.780 & 3.586 & 0.493 & 55.020 & 36.9 & 11.10 \\
Terminal-Bench 2.1 & RouteLLM & 150 & 2,330 & 36.852 & 29.530 & 3.272 & 1.230 & 70.883 & 44.5 & 21.66 \\
Terminal-Bench 2.1 & AutoMix & 159 & 2,178 & 35.608 & 24.130 & 3.821 & 0.822 & 66.480 & 40.4 & 16.51 \\
Terminal-Bench 2.1 & Plan & 150 & 2,450 & 30.332 & 25.749 & 3.932 & 0.907 & 60.920 & 45.9 & 11.34 \\
Terminal-Bench 2.1 & Heuristic & 150 & 3,106 & 55.478 & 55.717 & 3.928 & 2.356 & 117.479 & 50.1 & 27.47 \\
Terminal-Bench 2.1 & Always & 150 & 3,373 & 59.327 & 50.774 & 4.183 & 1.979 & 116.263 & 46.1 & 27.94 \\
Terminal-Bench 2.1 & LLM-Router & 150 & 2,809 & 49.815 & 43.169 & 4.947 & 1.124 & 99.055 & 46.4 & 18.47 \\
Terminal-Bench 2.1 & Random & 150 & 2,870 & 48.688 & 42.470 & 3.571 & 1.490 & 96.219 & 46.6 & 22.25 \\
Terminal-Bench 2.1 & ARIA & 150 & 3,356 & 51.080 & 57.531 & 3.811 & 1.993 & 114.415 & 53.0 & 24.93 \\
\midrule
\(\tau^2\)-bench Telecom & Executor-only & 120 & 4,852 & 21.605 & 25.813 & 0.978 & 0.058 & 48.455 & 54.4 & 5.07 \\
\(\tau^2\)-bench Telecom & Advisor-only & 120 & 4,288 & 18.024 & 22.365 & 0.785 & 1.462 & 42.635 & 55.4 & 22.50 \\
\(\tau^2\)-bench Telecom & FrugalGPT & 127 & 5,232 & 23.148 & 28.187 & 1.048 & 0.179 & 52.563 & 54.9 & 7.02 \\
\(\tau^2\)-bench Telecom & RouteLLM & 120 & 4,780 & 20.128 & 27.370 & 0.953 & 0.444 & 48.895 & 57.6 & 9.61 \\
\(\tau^2\)-bench Telecom & AutoMix & 218 & 8,665 & 37.079 & 47.209 & 1.688 & 1.312 & 88.611 & 56.0 & 24.98 \\
\(\tau^2\)-bench Telecom & Plan & 120 & 4,771 & 21.710 & 25.462 & 0.952 & 0.122 & 48.245 & 54.0 & 5.69 \\
\(\tau^2\)-bench Telecom & Heuristic & 120 & 4,764 & 21.926 & 25.933 & 0.933 & 0.268 & 49.059 & 54.2 & 7.13 \\
\(\tau^2\)-bench Telecom & Always & 120 & 5,121 & 23.072 & 29.640 & 0.991 & 0.458 & 54.162 & 56.2 & 9.01 \\
\(\tau^2\)-bench Telecom & LLM-Router & 120 & 5,106 & 23.877 & 28.333 & 1.037 & 0.101 & 53.348 & 54.3 & 6.16 \\
\(\tau^2\)-bench Telecom & Random & 120 & 4,905 & 21.537 & 28.631 & 0.958 & 0.478 & 51.604 & 57.1 & 8.24 \\
\(\tau^2\)-bench Telecom & ARIA & 120 & 4,897 & 21.359 & 28.469 & 0.955 & 0.434 & 51.217 & 57.1 & 7.77 \\
\bottomrule
\end{tabular}

    }
    \caption{Token usage and realized model cost for all headline evaluation
    conditions. Token columns are totals across three evaluation seeds in
    millions of tokens. ``Input'' denotes fresh input excluding cached prompt
    tokens; ``Cached \%'' is the cached share of total prompt tokens.
    ``Cost/run'' is the mean realized cost across the three evaluation seeds,
    corresponding to Table~2. For cascades, the episode column counts billed
    component executions rather than unique benchmark tasks.}
    \label{tab:token-summary}
\end{table}

\paragraph{Cost reconciliation.}
The reconstructed accounting reproduces all 33 headline cost cells. The
maximum absolute difference between the reconstructed mean and the value
printed in Table~2 is \$0.0049, and the maximum deviation from the signed-off
per-seed costs is \$0.0006. The per-call pricing equation itself reproduces
100\% of the 106{,}662 non-failed priced calls for which complete token records
are available.

\paragraph{Accounting completeness.}
A small number of episodes lack complete per-call token records. In these
cases, reported costs are retained from the episode-level ledgers rather than
inferred from token counts. Consequently, condition-level costs remain
unchanged, while token aggregates include only recorded per-call usage.

\clearpage
\section{Additional Results and Analyses}
\label{app:additional-results}

\subsection{Seed-Level Dispersion}
\label{app:dispersion}

To quantify run-to-run variation, Table~\ref{tab:seed-dispersion} reports
sample standard deviations across the three evaluation seeds corresponding to
the means in Table~2. These values describe dispersion across repeated runs
rather than confidence intervals over the underlying task distribution.

Variation is not specific to \app. Its success standard deviation is
3.5, 5.0, and 2.9 percentage points on SWE-bench, Terminal-Bench, and
$\tau^2$-bench, respectively, while several comparison conditions exhibit
equal or larger variation. We therefore interpret small differences between
nearby operating points descriptively rather than as evidence of statistical
separation.

\begin{table}[H]
    \centering
    \small
    \caption{Task success and realized model cost across three evaluation
    seeds. Entries report mean $\pm$ sample standard deviation ($n=3$).}
    \label{tab:seed-dispersion}
    \begin{tabular}{lcc cc cc}
        \toprule
        & \multicolumn{2}{c}{\textbf{SWE-bench Verified}}
        & \multicolumn{2}{c}{\textbf{Terminal-Bench 2.1}}
        & \multicolumn{2}{c}{$\boldsymbol{\tau^2}$\textbf{-bench Telecom}} \\
        \cmidrule(lr){2-3}
        \cmidrule(lr){4-5}
        \cmidrule(lr){6-7}
        Condition
        & Succ.\ (\%) & Cost (\$)
        & Succ.\ (\%) & Cost (\$)
        & Succ.\ (\%) & Cost (\$) \\
        \midrule

        Executor-only
        & $18.0 \pm 4.0$ & $11.48 \pm 2.30$
        & $21.3 \pm 3.1$ & $9.76 \pm 1.90$
        & $72.5 \pm 4.3$ & $5.07 \pm 0.04$ \\

        Advisor-only
        & $62.0 \pm 7.2$ & $45.90 \pm 5.30$
        & $51.3 \pm 1.2$ & $60.99 \pm 5.05$
        & $98.3 \pm 2.9$ & $22.50 \pm 1.22$ \\

        FrugalGPT
        & $31.3 \pm 2.3$ & $22.61 \pm 1.44$
        & $22.0 \pm 4.0$ & $11.10 \pm 2.89$
        & $78.3 \pm 2.9$ & $7.02 \pm 0.40$ \\

        RouteLLM
        & $30.0 \pm 5.3$ & $25.03 \pm 2.35$
        & $28.0 \pm 3.5$ & $21.66 \pm 0.85$
        & $88.3 \pm 3.8$ & $9.61 \pm 0.12$ \\

        AutoMix
        & $52.0 \pm 3.5$ & $59.68 \pm 5.75$
        & $22.0 \pm 2.0$ & $16.51 \pm 4.21$
        & $97.5 \pm 2.5$ & $24.98 \pm 0.64$ \\

        Plan
        & $24.0 \pm 7.2$ & $10.90 \pm 0.42$
        & $24.0 \pm 5.3$ & $11.34 \pm 0.67$
        & $70.0 \pm 4.3$ & $5.69 \pm 0.27$ \\

        Heuristic
        & $41.3 \pm 3.1$ & $21.50 \pm 1.35$
        & $34.7 \pm 3.1$ & $27.47 \pm 2.76$
        & $83.3 \pm 5.2$ & $7.13 \pm 0.10$ \\

        Always
        & $34.0 \pm 2.0$ & $21.39 \pm 0.88$
        & $35.3 \pm 4.2$ & $27.94 \pm 2.68$
        & $87.5 \pm 2.5$ & $9.01 \pm 0.12$ \\

        LLM-Router
        & $29.3 \pm 4.2$ & $15.77 \pm 1.76$
        & $30.0 \pm 2.0$ & $18.47 \pm 0.96$
        & $87.5 \pm 4.3$ & $6.16 \pm 0.34$ \\

        \textit{Random}
        & $28.0 \pm 2.0$ & $17.33 \pm 2.11$
        & $30.7 \pm 5.0$ & $22.25 \pm 3.07$
        & $85.8 \pm 5.2$ & $8.24 \pm 0.51$ \\

        \textit{\app}
        & $48.0 \pm 3.5$ & $22.09 \pm 2.65$
        & $34.7 \pm 5.0$ & $24.93 \pm 2.73$
        & $93.3 \pm 2.9$ & $7.77 \pm 1.47$ \\

        \bottomrule
    \end{tabular}
\end{table}

\subsection{Placement Analysis}
\label{app:placements}

\begin{figure}[H]
    \centering
    \includegraphics[
        width=0.9\linewidth,
        trim=0 21 0 23,
        clip
    ]{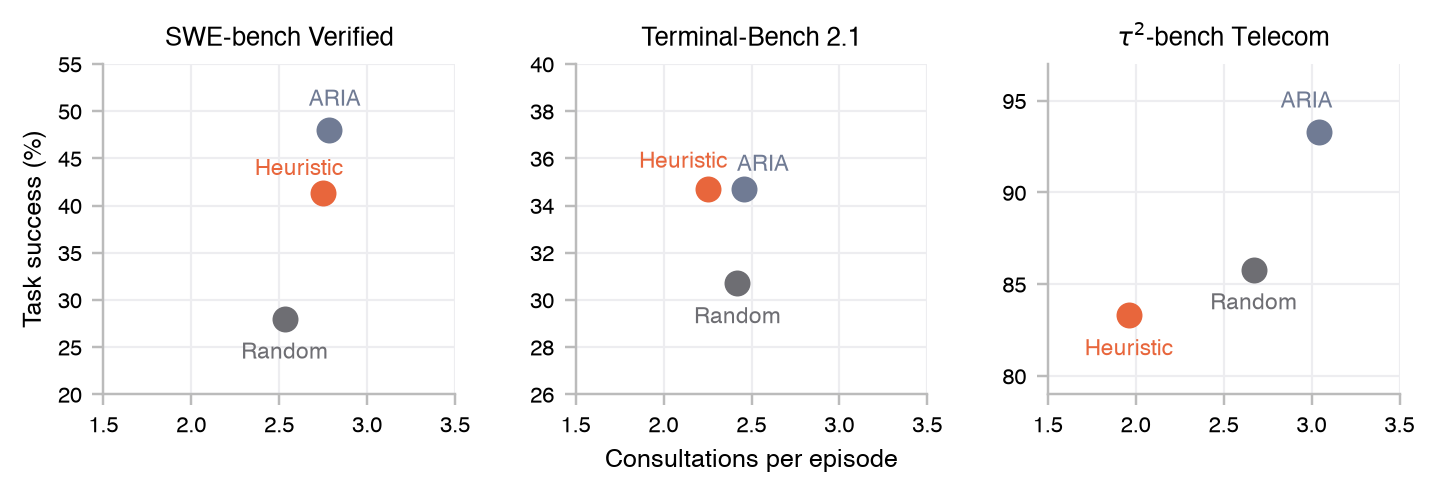}
    \caption{Task success and consultation frequency for \app, \textit{Heuristic}, and \textit{Random} across all three benchmarks.}
    \label{fig:placement}
\end{figure}

Section~5.1 compares \app{} with \textit{Heuristic} and \textit{Random} to
separate consultation placement from advisor access and consultation volume.
Figure~\ref{fig:placement} visualizes this comparison by plotting task success
against the realized number of consultations per episode. On SWE-bench, the
three conditions operate at similar consultation volumes but occupy clearly
different success levels, with \app{} above both fixed and uninformed
placement. The separation is smaller on Terminal-Bench, where \app{} and
\textit{Heuristic} reach the same mean success despite allocating advice
differently. On $\tau^2$-bench, \app{} again attains higher success than both
comparison conditions, although \textit{Random} realizes somewhat less advice
than the target volume. The figure therefore complements the aggregate
success--cost results by showing how consultation outcomes vary with where,
rather than only how often, advisor access is allocated.

\subsection{Training Dynamics and Policy Inspection}
\label{app:policy-diagnostics}

We examine the controller through two complementary views: its behavior
during training and the action values underlying its frozen decisions.

\begin{figure}[htbp]
    \centering
    \includegraphics[width=\linewidth]{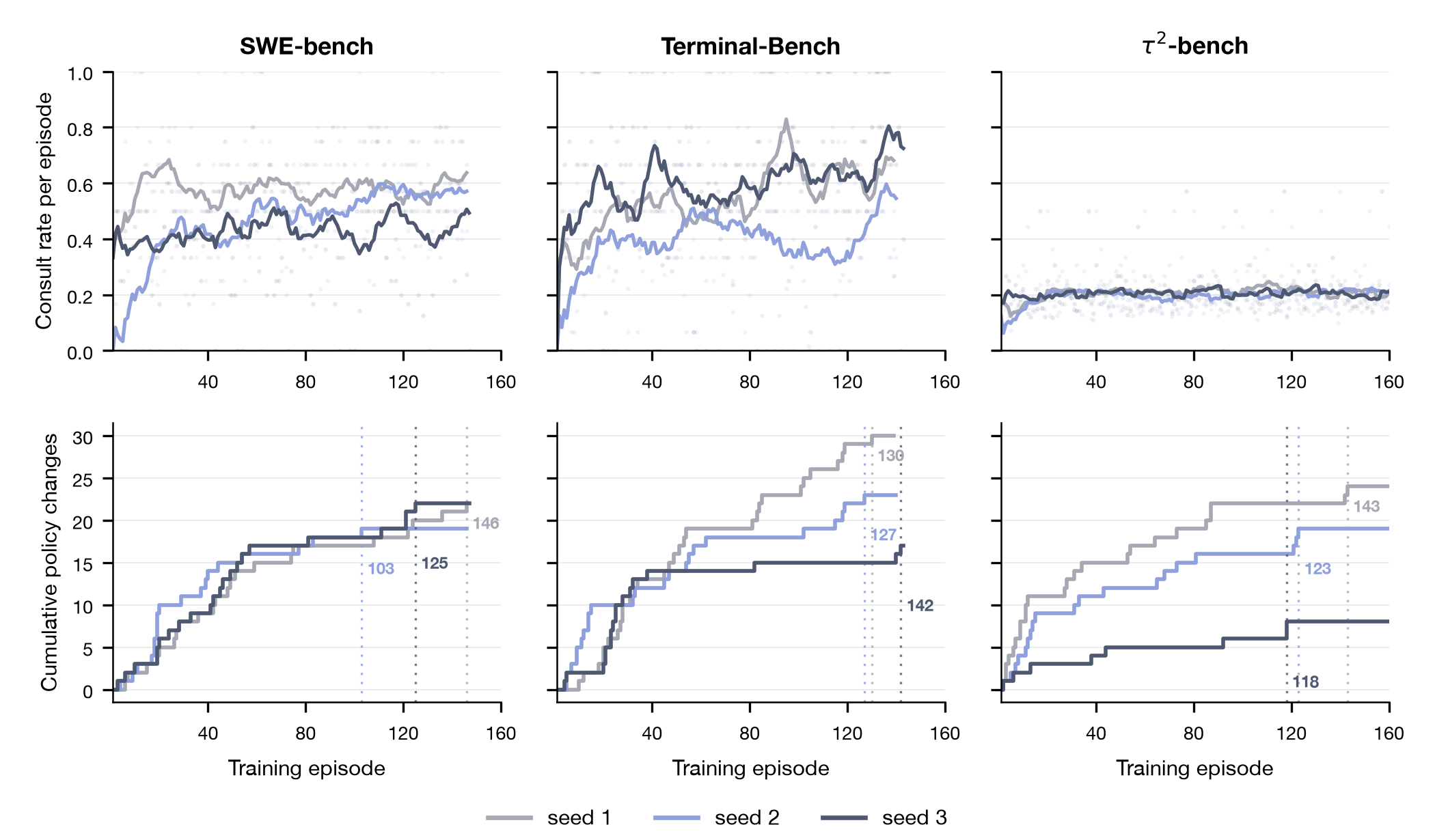}
    \caption{Training dynamics across three benchmarks and three
    controller seeds.
    \textbf{Top:} per-episode consultation rates (faint points) and
    smoothed trajectories (solid lines).
    \textbf{Bottom:} cumulative changes in the greedy tabular policy.
    Dotted vertical lines mark the last observed greedy-policy change
    for each seed, not a certified convergence point.}
    \label{fig:training-dynamics}
\end{figure}

\paragraph{Training dynamics.}
Figure~\ref{fig:training-dynamics} compares consultation frequency with
changes in the greedy policy across the three training seeds of each
benchmark. On $\tau^2$-bench, consultation rates remain comparatively
similar across seeds even while individual greedy decisions continue
to change. SWE-bench and Terminal-Bench show greater variation in
consultation frequency. The two views therefore capture different
properties: a stable aggregate consultation rate does not imply an
unchanged state-conditioned policy. These plots describe observed
training behavior rather than establish policy convergence.

\paragraph{Inspecting the frozen policies.}
Figures~\ref{fig:qvalues-swe}--\ref{fig:qvalues-tau} report the learned
action values and state visit counts for each benchmark and controller seed,
together with corresponding heatmap visualizations. The $Q$-values are
empirical estimates of cost-aware return rather than calibrated success
probabilities. For each state, the greedy policy compares
$Q(s,\textsc{Consult})$ with $Q(s,\textsc{Silent})$; the larger value
determines the preferred action, subject to the tie-breaking rule and budget
constraint in Appendix~\ref{app:training}. The relevant comparison is therefore
between actions within the same state rather than between absolute values
across benchmarks or seeds.

The compact tabular representation makes each consultation decision directly
inspectable, while the heatmaps expose broader allocation patterns.
Table~\ref{tab:policy-consensus} complements these per-seed policies by
summarizing agreement in their resulting greedy actions across seeds.


\newcommand{\SeedTableBlock}[2]{%
    \noindent
    %
    \begin{minipage}[c]{0.06\linewidth}
        \centering
        \rotatebox[origin=c]{90}{%
            \scriptsize\bfseries #1%
        }%
    \end{minipage}%
    \hspace{0.01\linewidth}%
    %
    \begin{minipage}[c]{0.91\linewidth}
        \centering

        {\fontsize{7.4}{7.9}\selectfont
        \setlength{\tabcolsep}{1.7pt}
        \renewcommand{\arraystretch}{0.92}

        \begin{tabular*}{\linewidth}{
            @{\extracolsep{\fill}}lrr@{}
        }
            \toprule
            State & $N$ & $Q_S/Q_C$ \\
            \midrule
            #2
            \bottomrule
        \end{tabular*}
        }

    \end{minipage}%

    \par\vspace{0.45em}
}


\newcommand{\SWESeedOneRows}{%
F/N/0 & 149 & 0.033/-0.007 \\
E/N/0 &  38 & 0.095/0.017 \\
E/N/1 & 171 & 0.041/0.011 \\
E/C/0 &   1 & 0.029/0.000 \\
E/C/1 &  22 & -0.075/0.109 \\
E/X/0 &   1 & 0.029/0.000 \\
E/X/1 &  22 & -0.116/0.269 \\
R/N/0 &  16 & 0.258/-0.228 \\
R/N/1 &  43 & -0.077/-0.112 \\
R/C/0 &   6 & -0.142/0.000 \\
R/C/1 &  86 & 0.152/0.010 \\
R/X/0 &   0 & 0.000/0.000 \\
R/X/1 &  47 & -0.266/-0.031 \\
D/C/0 &   9 & -0.912/0.311 \\
D/C/1 & 129 & -0.100/0.095 \\
D/X/0 &   8 & 0.000/-0.044 \\
D/X/1 &  81 & -0.028/0.159 \\
}


\newcommand{\SWESeedTwoRows}{%
F/N/0 & 149 & 0.007/-0.055 \\
E/N/0 & 113 & -0.065/0.029 \\
E/N/1 &  97 & 0.028/0.009 \\
E/C/0 &   8 & -0.426/0.091 \\
E/C/1 &  12 & -0.279/0.168 \\
E/X/0 &   5 & 0.418/0.000 \\
E/X/1 &  15 & 0.169/-0.045 \\
R/N/0 &   1 & 0.088/0.000 \\
R/N/1 &  44 & 0.064/-0.096 \\
R/C/0 &   1 & -0.882/0.000 \\
R/C/1 &  88 & 0.049/-0.108 \\
R/X/0 &   0 & 0.000/0.000 \\
R/X/1 &  68 & -0.277/0.148 \\
D/C/0 &  16 & -0.338/0.224 \\
D/C/1 & 115 & 0.088/0.129 \\
D/X/0 &  21 & -0.533/-0.017 \\
D/X/1 &  88 & -0.079/0.011 \\
}


\newcommand{\SWESeedThreeRows}{%
F/N/0 & 149 & 0.014/-0.147 \\
E/N/0 & 173 & 0.047/-0.135 \\
E/N/1 &  35 & -0.093/-0.170 \\
E/C/0 &  14 & -0.243/-0.178 \\
E/C/1 &  12 & 0.055/-0.311 \\
E/X/0 &   5 & -0.515/0.375 \\
E/X/1 &  14 & 0.467/0.153 \\
R/N/0 &  35 & 0.078/-0.431 \\
R/N/1 &  16 & -0.213/-0.412 \\
R/C/0 &  37 & 0.125/0.157 \\
R/C/1 &  86 & -0.101/0.042 \\
R/X/0 &   5 & -0.022/-1.000 \\
R/X/1 &  39 & -0.103/0.127 \\
D/C/0 &  41 & -0.279/0.110 \\
D/C/1 &  81 & -0.118/0.036 \\
D/X/0 &  48 & -0.434/0.192 \\
D/X/1 &  57 & -0.179/0.037 \\
}


\clearpage

\begin{figure}[p]
    \centering

    \begin{minipage}[t]{0.34\textwidth}
        \vspace{2.2em}

        \SeedTableBlock{Seed 1}{\SWESeedOneRows}

        \SeedTableBlock{Seed 2}{\SWESeedTwoRows}

        \SeedTableBlock{Seed 3}{\SWESeedThreeRows}

    \end{minipage}
    \hfill
    \begin{minipage}[t]{0.63\textwidth}
        \vspace{0pt}
        \centering

        \includegraphics[
            width=0.99\linewidth,
            trim=26 0 6 16,
            clip
        ]{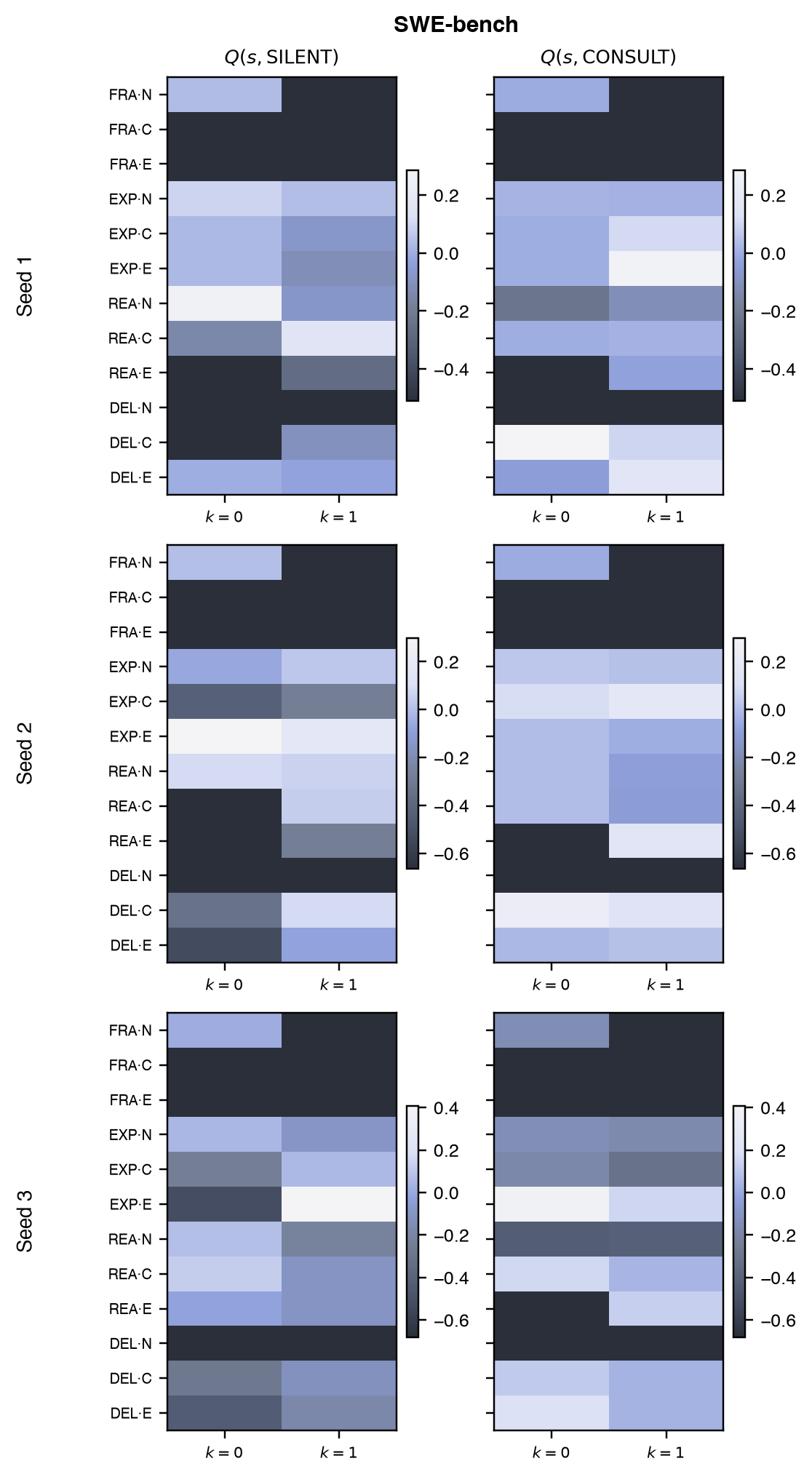}

    \end{minipage}

    \caption{
        Learned action values for SWE-bench Verified across three independently
        trained controller seeds.
        \textbf{Left:} compact numerical tables reporting state visit counts
        and action values, where
        $Q_S = Q(s,\textsc{Silent})$ and
        $Q_C = Q(s,\textsc{Consult})$.
        State notation is \emph{phase/evidence/prior consultation}:
        F/E/R/D denote \textsc{Frame}, \textsc{Explore}, \textsc{Reason}, and
        \textsc{Deliver}; N/C/X denote \textsc{None}, \textsc{Clean}, and
        \textsc{Error}. Structurally unreachable states are omitted, while
        reachable states with $N=0$ are retained.
        \textbf{Right:} the corresponding learned value tables for the three
        seeds visualized as heatmaps. Deployment selects the action with the
        larger value within each state; ties resolve to \textsc{Silent}.
    }
    \label{fig:qvalues-swe}

\end{figure}

\clearpage

\newcommand{\TBSeedOneRows}{%
F/N/0 & 139 & 0.246/0.434 \\
E/N/0 &   2 & 1.000/0.000 \\
E/N/1 &   2 & -0.073/0.000 \\
E/C/0 &   6 & 0.476/-0.049 \\
E/C/1 &  90 & 0.342/0.107 \\
E/X/0 &   2 & -0.049/-0.049 \\
E/X/1 &  45 & 0.331/0.341 \\
R/N/0 &   1 & 1.000/0.000 \\
R/N/1 &  57 & -0.013/-0.049 \\
R/C/0 &   1 & 1.000/0.000 \\
R/C/1 & 188 & 0.028/0.242 \\
R/X/0 &   0 & 0.000/0.000 \\
R/X/1 & 105 & 0.002/0.224 \\
D/C/0 &   5 & 0.750/-0.049 \\
D/C/1 & 137 & 0.484/0.512 \\
D/X/0 &   1 & 0.000/0.000 \\
D/X/1 &  34 & 0.000/0.236 \\
}


\newcommand{\TBSeedTwoRows}{%
F/N/0 & 140 & 0.483/0.257 \\
E/N/0 &   0 & 0.000/0.000 \\
E/N/1 &   0 & 0.000/0.000 \\
E/C/0 &  68 & 0.309/0.417 \\
E/C/1 &  43 & 0.330/0.151 \\
E/X/0 &  23 & 0.368/0.186 \\
E/X/1 &  13 & 0.142/0.237 \\
R/N/0 &   0 & 0.000/0.000 \\
R/N/1 &   0 & 0.000/0.000 \\
R/C/0 &   1 & 0.000/0.000 \\
R/C/1 & 413 & 0.043/0.007 \\
R/X/0 &   1 & -0.024/0.000 \\
R/X/1 &  28 & -0.022/-0.058 \\
D/C/0 &  45 & 0.625/0.722 \\
D/C/1 & 109 & 0.286/0.467 \\
D/X/0 &  18 & 0.167/0.460 \\
D/X/1 &  20 & 0.143/-0.032 \\
}


\newcommand{\TBSeedThreeRows}{%
F/N/0 & 143 & 0.537/0.540 \\
E/N/0 &   0 & 0.000/0.000 \\
E/N/1 &   5 & -0.073/-0.073 \\
E/C/0 &  11 & 0.339/0.000 \\
E/C/1 &  89 & 0.495/0.393 \\
E/X/0 &   9 & 0.000/0.641 \\
E/X/1 &  40 & 0.619/0.512 \\
R/N/0 &   0 & 0.000/0.000 \\
R/N/1 &  57 & -0.026/-0.049 \\
R/C/0 &   0 & 0.000/0.000 \\
R/C/1 & 177 & 0.043/0.253 \\
R/X/0 &   0 & 0.000/0.000 \\
R/X/1 &  49 & -0.007/0.191 \\
D/C/0 &   9 & 0.667/0.594 \\
D/C/1 & 152 & 0.290/0.700 \\
D/X/0 &   5 & 0.200/0.000 \\
D/X/1 &  27 & 0.071/0.409 \\
}


\clearpage

\begin{figure}[p]
    \centering

    \begin{minipage}[t]{0.34\textwidth}
        \vspace{2.2em}

        \SeedTableBlock{Seed 1}{\TBSeedOneRows}

        \SeedTableBlock{Seed 2}{\TBSeedTwoRows}

        \SeedTableBlock{Seed 3}{\TBSeedThreeRows}

    \end{minipage}
    \hfill
    \begin{minipage}[t]{0.63\textwidth}
        \vspace{0pt}
        \centering

        \includegraphics[
            width=0.99\linewidth,
            trim=26 0 6 16,
            clip
        ]{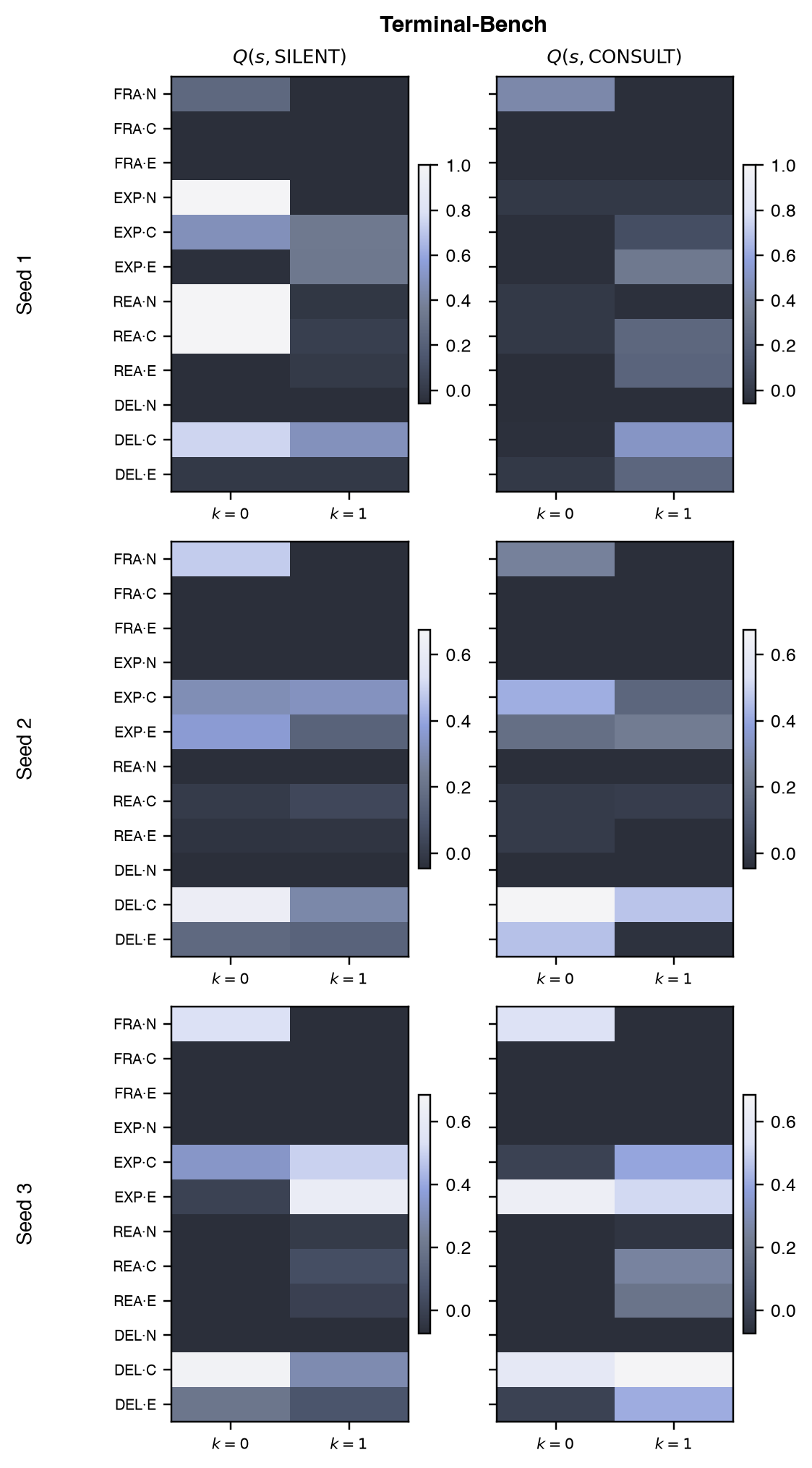}

    \end{minipage}

    \caption{
        Learned action values for Terminal-Bench 2.1 across three independently
        trained controller seeds.
        \textbf{Left:} compact numerical tables reporting state visit counts
        and action values, where
        $Q_S = Q(s,\textsc{Silent})$ and
        $Q_C = Q(s,\textsc{Consult})$.
        State notation is \emph{phase/evidence/prior consultation}:
        F/E/R/D denote \textsc{Frame}, \textsc{Explore}, \textsc{Reason}, and
        \textsc{Deliver}; N/C/X denote \textsc{None}, \textsc{Clean}, and
        \textsc{Error}. Structurally unreachable states are omitted, while
        reachable states with $N=0$ are retained.
        \textbf{Right:} the corresponding learned value tables for the three
        seeds visualized as heatmaps. Deployment selects the action with the
        larger value within each state; ties resolve to \textsc{Silent}.
    }
    \label{fig:qvalues-tb}

\end{figure}

\clearpage


\newcommand{\TauSeedOneRows}{%
F/N/0 & 160  & 0.733/0.555 \\
E/N/0 & 167  & 0.673/0.757 \\
E/N/1 &  74  & 0.647/0.543 \\
E/O/0 &  27  & 0.674/0.750 \\
E/O/1 &  52  & 0.745/0.848 \\
E/F/0 &   0  & 0.000/0.000 \\
E/F/1 &   0  & 0.000/0.000 \\
R/N/0 &   1  & -0.076/0.000 \\
R/N/1 &   4  & 0.943/0.000 \\
R/O/0 &  51  & 0.393/0.638 \\
R/O/1 & 2641 & 0.745/0.749 \\
R/F/0 &   0  & 0.000/0.000 \\
R/F/1 &   0  & 0.000/0.000 \\
D/O/0 &   1  & 0.000/0.000 \\
D/O/1 & 159  & 0.792/0.000 \\
D/F/0 &   0  & 0.000/0.000 \\
D/F/1 &   0  & 0.000/0.000 \\
}


\newcommand{\TauSeedTwoRows}{%
F/N/0 & 160  & 0.512/0.690 \\
E/N/0 &  30  & 0.465/0.433 \\
E/N/1 & 223  & 0.701/0.678 \\
E/O/0 &   7  & 0.803/0.000 \\
E/O/1 &  60  & 0.779/0.722 \\
E/F/0 &   0  & 0.000/0.000 \\
E/F/1 &   0  & 0.000/0.000 \\
R/N/0 &   1  & 0.924/0.000 \\
R/N/1 &   5  & 0.943/0.000 \\
R/O/0 &  30  & 0.137/0.554 \\
R/O/1 & 2711 & 0.694/0.708 \\
R/F/0 &   0  & 0.000/0.000 \\
R/F/1 &   4  & 1.000/0.000 \\
D/O/0 &   1  & 0.000/0.000 \\
D/O/1 & 158  & 0.740/0.856 \\
D/F/0 &   0  & 0.000/0.000 \\
D/F/1 &   1  & 1.000/0.000 \\
}


\newcommand{\TauSeedThreeRows}{%
F/N/0 & 160  & 0.599/0.745 \\
E/N/0 &  21  & 0.568/0.424 \\
E/N/1 & 223  & 0.769/0.415 \\
E/O/0 &   7  & 0.760/0.924 \\
E/O/1 &  69  & 0.839/0.610 \\
E/F/0 &   0  & 0.000/0.000 \\
E/F/1 &   0  & 0.000/0.000 \\
R/N/0 &   2  & 0.943/0.000 \\
R/N/1 &  11  & -0.038/0.843 \\
R/O/0 &  91  & 0.662/0.656 \\
R/O/1 & 2624 & 0.747/0.784 \\
R/F/0 &   0  & 0.000/0.000 \\
R/F/1 &   0  & 0.000/0.000 \\
D/O/0 &   1  & 0.000/0.000 \\
D/O/1 & 159  & 0.809/0.981 \\
D/F/0 &   0  & 0.000/0.000 \\
D/F/1 &   0  & 0.000/0.000 \\
}


\clearpage

\begin{figure}[p]
    \centering

    \begin{minipage}[t]{0.34\textwidth}
        \vspace{2.2em}

        \SeedTableBlock{Seed 1}{\TauSeedOneRows}

        \SeedTableBlock{Seed 2}{\TauSeedTwoRows}

        \SeedTableBlock{Seed 3}{\TauSeedThreeRows}

    \end{minipage}
    \hfill
    \begin{minipage}[t]{0.63\textwidth}
        \vspace{0pt}
        \centering

        \includegraphics[
            width=0.99\linewidth,
            trim=26 0 6 16,
            clip
        ]{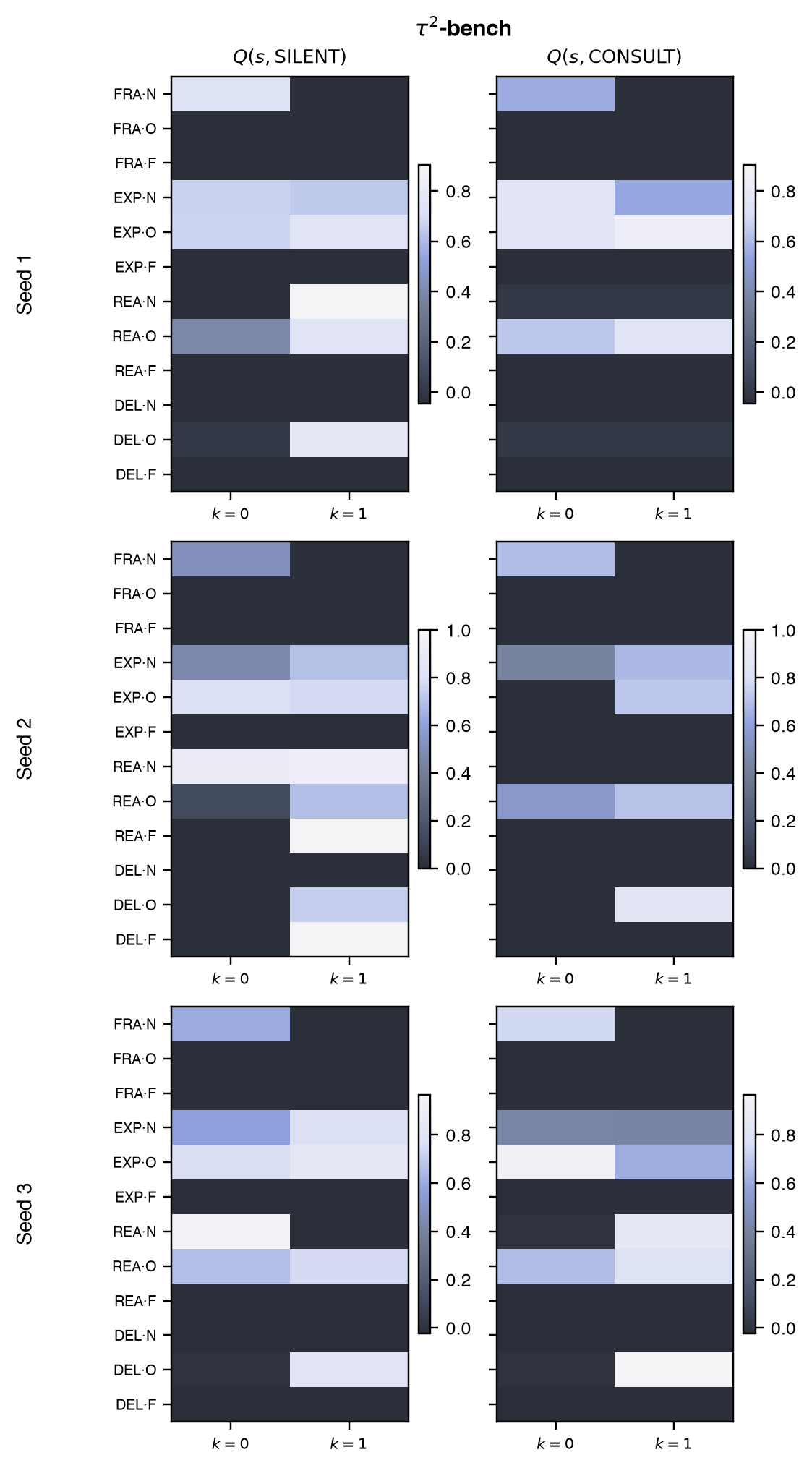}

    \end{minipage}

    \caption{
        Learned action values for $\tau^2$-bench Telecom across three
        independently trained controller seeds.
        \textbf{Left:} compact numerical tables reporting state visit counts
        and action values, where
        $Q_S = Q(s,\textsc{Silent})$ and
        $Q_C = Q(s,\textsc{Consult})$.
        State notation is \emph{phase/evidence/prior consultation}:
        F/E/R/D denote \textsc{Frame}, \textsc{Explore}, \textsc{Reason}, and
        \textsc{Deliver}; N/O/F denote \textsc{None}, \textsc{OK}, and
        \textsc{Failed}. Structurally unreachable states are omitted, while
        reachable states with $N=0$ are retained.
        \textbf{Right:} the corresponding learned value tables for the three
        seeds visualized as heatmaps. Deployment selects the action with the
        larger value within each state; ties resolve to \textsc{Silent}.
    }
    \label{fig:qvalues-tau}

\end{figure}

\clearpage

\subsection{Policy Consensus}
\label{app:policy_consensus}

Table~\ref{tab:policy-consensus} reports the greedy consultation decisions learned independently across the three seeds of each benchmark. The All column aggregates all nine frozen policies for direct cross-benchmark comparison; unvisited states contribute the deployment default \textsc{SILENT}. Agreement is limited overall, reflecting benchmark-specific allocation, but one state recurs across nearly all policies: at \textsc{Deliver} with clean/OK evidence after prior consultation, eight of nine frozen tables select \textsc{Consult}.

\begin{table}[H]
    \centering
    \small
    \caption{
        Cross-seed agreement of the learned greedy consultation policies.
        Entries report the number of seeds whose frozen table selects
        \textsc{Consult} in a given state. A dash indicates that the state was
        never visited during training for that benchmark. For $\tau^2$,
        \textsc{OK} and \textsc{Failed} correspond to the
        \textsc{Clean} and \textsc{Error} evidence classes used on
        SWE-bench and Terminal-Bench.
    }
    \label{tab:policy-consensus}
    \setlength{\tabcolsep}{5pt}
    \renewcommand{\arraystretch}{1.05}

    \begin{tabular}{lllcccc}
        \toprule
        \textbf{Position}
        & \textbf{Evidence}
        & \textbf{Prior consult}
        & \textbf{SWE}
        & \textbf{TB}
        & $\boldsymbol{\tau^2}$
        & \textbf{All} \\
        \midrule

        Frame
        & None
        & No
        & 0/3 & 2/3 & 2/3 & 4/9 \\
        \midrule

        \multirow{6}{*}{Explore}
        & None       & No  & 1/3 & 0/3 & 1/3 & 2/9 \\
        & None       & Yes & 0/3 & 1/3 & 0/3 & 1/9 \\
        & Clean/OK   & No  & 2/3 & 1/3 & 2/3 & 5/9 \\
        & Clean/OK   & Yes & 2/3 & 0/3 & 1/3 & 3/9 \\
        & Error/Fail & No  & 1/3 & 1/3 & --  & 2/9 \\
        & Error/Fail & Yes & 1/3 & 2/3 & --  & 3/9 \\
        \midrule

        \multirow{6}{*}{Reason}
        & None       & No  & 0/3 & 0/3 & 1/3 & 1/9 \\
        & None       & Yes & 0/3 & 0/3 & 1/3 & 1/9 \\
        & Clean/OK   & No  & 3/3 & 0/3 & 2/3 & 5/9 \\
        & Clean/OK   & Yes & 1/3 & 2/3 & 3/3 & 6/9 \\
        & Error/Fail & No  & 0/3 & 1/3 & --  & 1/9 \\
        & Error/Fail & Yes & 3/3 & 2/3 & 0/3 & 5/9 \\
        \midrule

        \multirow{4}{*}{Deliver}
        & Clean/OK   & No
        & 3/3 & 1/3 & 0/3 & 4/9 \\

        & \textbf{Clean/OK}
        & \textbf{Yes}
        & \textbf{3/3}
        & \textbf{3/3}
        & \textbf{2/3}
        & \textbf{8/9} \\

        & Error/Fail & No
        & 2/3 & 1/3 & -- & 3/9 \\

        & Error/Fail & Yes
        & 3/3 & 2/3 & 0/3 & 5/9 \\
        \bottomrule
    \end{tabular}
\end{table}

\subsection{Within-State Consultation Effects}
\label{app:ipw}

The learned value table estimates consultation value from the remaining
episodic return. As a complementary analysis, we use the randomized exploration
during controller training to estimate how consultation changes outcomes within
individual state cells. This analysis is distinct from the consultation-price
sweep in Section~5.2: the price sweep re-estimates a policy from the same
experience under a different reward penalty, whereas the analysis below
compares \textsc{Consult} and \textsc{Silent} decisions taken within the same
abstract state.

\paragraph{Estimator.}
During training, the $\epsilon$-greedy behavior policy logs the probability of
consultation for every decision. Because $\epsilon$ is bounded below by $0.10$,
the minority action has probability at least $0.05$, bounding the largest
inverse-propensity weight at $20$. Decisions forced by initialization or by an
exhausted consultation budget contain no randomized action choice and are
excluded.

For each state cell $s$, we estimate the self-normalized H\'ajek contrast
\[
\hat{\tau}_s =
\frac{\sum_{i:a_i=1} w_iY_i}
     {\sum_{i:a_i=1} w_i}
-
\frac{\sum_{i:a_i=0} w_iY_i}
     {\sum_{i:a_i=0} w_i},
\qquad
w_i =
\begin{cases}
1/p_i, & a_i=\textsc{Consult},\\
1/(1-p_i), & a_i=\textsc{Silent},
\end{cases}
\]
where $p_i$ is the logged consultation propensity and $Y_i$ the corresponding
outcome. No weight trimming is applied. Uncertainty is estimated with
10{,}000 bootstrap replicates resampling complete episodes rather than
individual decisions, preserving dependence among decisions from the same
trajectory. Cells with fewer than five observations for either action are not
assigned an effect estimate.

\begin{figure}[H]
    \centering
    \includegraphics[width=\linewidth]{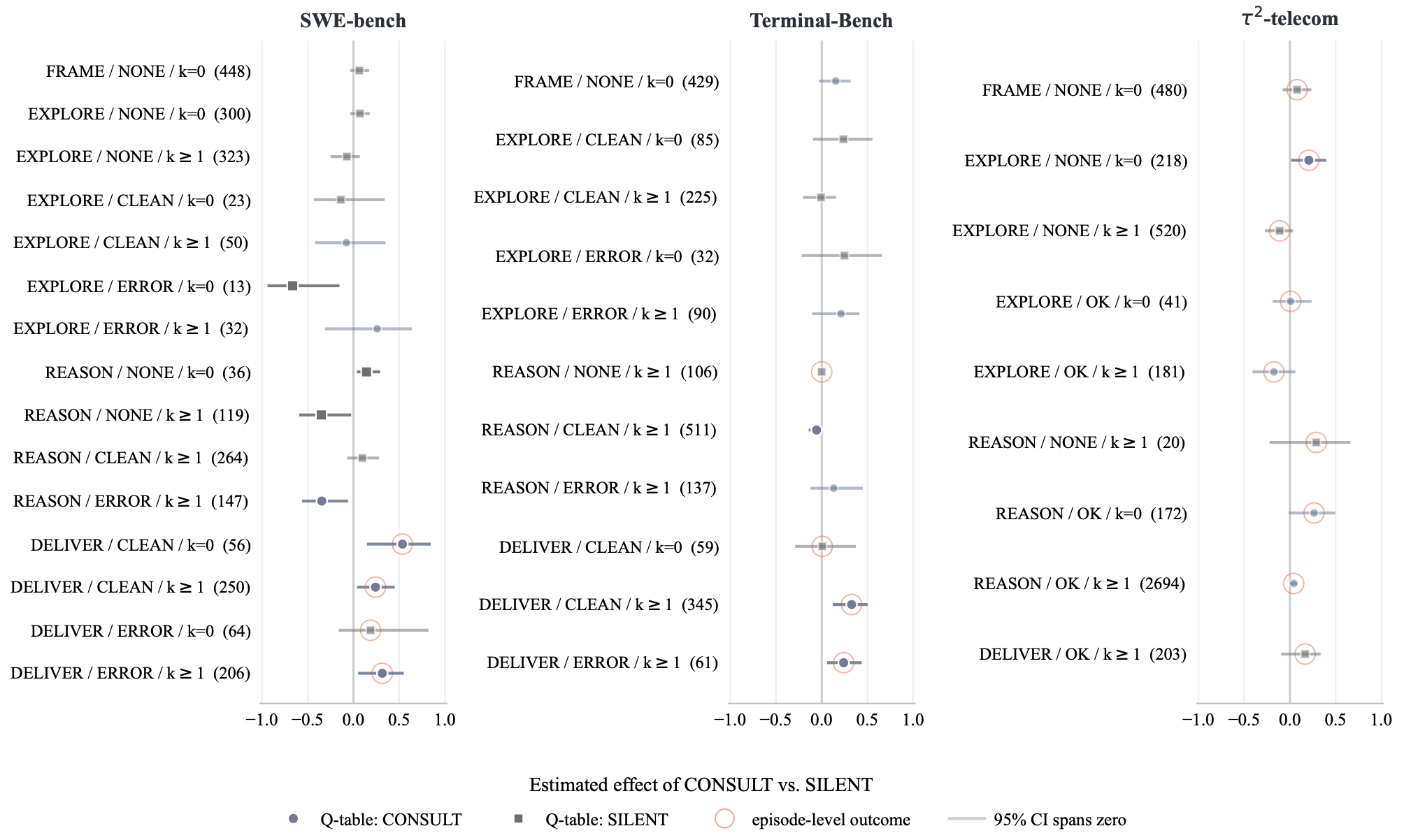}
    \caption{Propensity-weighted effect of \textsc{Consult} versus
    \textsc{Silent} within each state cell. Points show H\'ajek IPW estimates
    with 95\% episode-cluster bootstrap intervals. Marker shape indicates the
    greedy action of the reference $Q$-table; circled markers use the
    episode-level outcome because the one-step outcome is constant or
    unavailable. Faded intervals include zero. Parentheses report the number
    of observed decisions in each cell.}
    \label{fig:ipw}
\end{figure}

After these exclusions, the analysis contains 2{,}373 decisions from 448
SWE-bench episodes, 2{,}115 decisions from 429 Terminal-Bench episodes, and
4{,}541 decisions from 480 $\tau^2$-telecom episodes. In
Figure~\ref{fig:ipw}, $k=0$ denotes no previous consultation and $k\geq1$
denotes at least one previous consultation, corresponding to
$h_t=0$ and $h_t=1$ in the state representation.

\paragraph{Outcome definition.}
The primary outcome is whether the workflow step gated by the decision executes
without an observed error. This one-step outcome is informative for intermediate
execution states but is constant at some contact points. In particular,
post-delivery execution produces no subsequent error-bearing step, and
$\tau^2$-telecom does not expose the same process-output channel used by
SWE-bench and Terminal-Bench. For these cells, we instead use episode success
as the outcome. The resulting estimates should therefore be interpreted
accordingly: delivery-gate estimates remain closely tied to the final decision,
whereas the $\tau^2$ estimates reflect episode-level association under the
training behavior distribution. The underlying evidence mapping is documented
in Appendix~\ref{app:state_construction}.

\paragraph{Results.}
Figure~\ref{fig:ipw} shows the propensity-weighted effects for all estimable
cells. The strongest and most consistent positive effects occur at delivery
gates. On SWE-bench, consultation at
\textsc{Deliver}/\textsc{Clean}/$k=0$ increases estimated episode success by
$0.54$ with a 95\% bootstrap interval of $[0.15,0.85]$. The corresponding
effects after prior consultation are $0.24$ $[0.04,0.45]$ for
\textsc{Deliver}/\textsc{Clean} and $0.32$ $[0.05,0.55]$ for
\textsc{Deliver}/\textsc{Error}. Terminal-Bench shows the same pattern:
\textsc{Deliver}/\textsc{Clean}/$k\geq1$ yields $0.33$ $[0.12,0.50]$ and
\textsc{Deliver}/\textsc{Error}/$k\geq1$ yields $0.24$ $[0.06,0.44]$.

Mid-trajectory effects are more heterogeneous. On SWE-bench,
\textsc{Reason}/\textsc{None}/$k\geq1$ and
\textsc{Reason}/\textsc{Error}/$k\geq1$ have negative one-step estimates of
$-0.35$ $[-0.59,-0.03]$ and $-0.35$ $[-0.56,-0.06]$, respectively. For the
latter cell, however, the corresponding episode-level estimate is positive
($0.23$ $[-0.03,0.46]$). This pattern is consistent with consultation causing
additional intermediate work before any benefit appears in the final task
outcome, illustrating the difference between immediate execution effects and
the longer-horizon return optimized by \app.

\paragraph{Relation to the learned policy.}
Among the eleven estimated cells whose 95\% interval excludes zero, the sign of
the propensity-weighted effect agrees with the greedy action of the reference
$Q$-table in eight; all five significant delivery-gate cells agree. The three
disagreements occur in intermediate \textsc{Reason} states where the
one-step estimate is negative while the learned policy selects
\textsc{Consult}. This is not necessarily contradictory: the $Q$-table is
trained on the remaining episodic return, whereas the primary IPW outcome
measures the next execution step. The agreement at delivery gates is therefore
the most direct comparison between the two estimands.

\paragraph{Limitations.}
The estimates use training trajectories because randomized exploration is
required for propensity weighting; the frozen evaluation policy is greedy.
They therefore characterize consultation effects under the observed training
distribution rather than under the deployed policy. Cell occupancy is also
highly uneven, and several cells are too sparse to estimate reliably.
For $\tau^2$-telecom, the analysis relies on episode success because the
one-step execution outcome is unavailable. Finally, the reported intervals are
not adjusted for multiple comparisons: 11 of 35 primary intervals exclude
zero, so individual cells should be interpreted together with the broader
cross-cell pattern rather than as isolated hypothesis tests.

\subsection{MAST Failure Analysis}
\label{app:mast}

\paragraph{Annotation protocol.}
We analyze failed evaluation trajectories using the 14-mode MAST taxonomy
\citep{cemri_2025_multiagent}, grouped into system design (FC1),
inter-agent misalignment (FC2), and task verification (FC3). Each trajectory
receives non-exclusive presence labels and one mutually exclusive primary
failure mode. Labels are produced by \texttt{gemini-3.1-pro-preview} at high
reasoning effort and temperature 0. Main-text comparisons use the primary
label; presence labels are used only as supporting diagnostics.

\paragraph{Coverage.}
MAST is applied only to failed trajectories with a usable execution trace;
successful trajectories are not judged. Episodes without a recoverable trace
remain failures in the headline benchmark metrics but are excluded from MAST
denominators. This affects Terminal-Bench in particular, where some crash-path
episodes contain no trajectory suitable for annotation. Across benchmarks,
over 1,600 failed trajectories are judged.

\paragraph{Human validation.}
Because the automatic judge belongs to the same model
family as the advisor, we manually validate a stratified sample of
40 trajectories. We record the primary MAST mode and, for
advisor conditions, separately determine whether relevant guidance was
available to the executor and whether it was reflected in subsequent actions.

On the primary label, human annotation agrees with the automatic judge on
35 out of 40 trajectories (87.5\%): 10/10 for both \textit{Executor-only} and
\textit{Always}, 8/10 for \textit{Advisor-only}, and 7/10 for \app. For context, MAST's original validation reports 94\% accuracy and Cohen's $\kappa = 0.77$ between its LLM annotator and expert annotations, providing a reference point for the 87.5\% exact primary-label agreement observed here \citep{cemri_2025_multiagent}. The dominant errors are  \emph{ignored agent input} (FM-2.5) and
\emph{incorrect verification} (FM-3.3). Additionally, manual validation confirms that in every advisor trajectory whose final
consult step flagged the run as off track, the relevant guidance had
reached the executor, yet the executor made no corrective action and declared the task as satisfied. Human validation thus supports
the claim that FM-2.5 reflects available advice that was not acted
upon, rather than advice that failed to reach the executor.

\paragraph{Interpretation.}
MAST labels characterize the composition of residual failures and are not
treated as causal attributions to individual consultations. Moreover, several
FC2 modes require inter-agent communication to occur at all; comparisons of
failure profiles should therefore be interpreted together with each
condition's communication regime.

\begin{figure}[H]
    \centering
    \includegraphics[width=\linewidth]{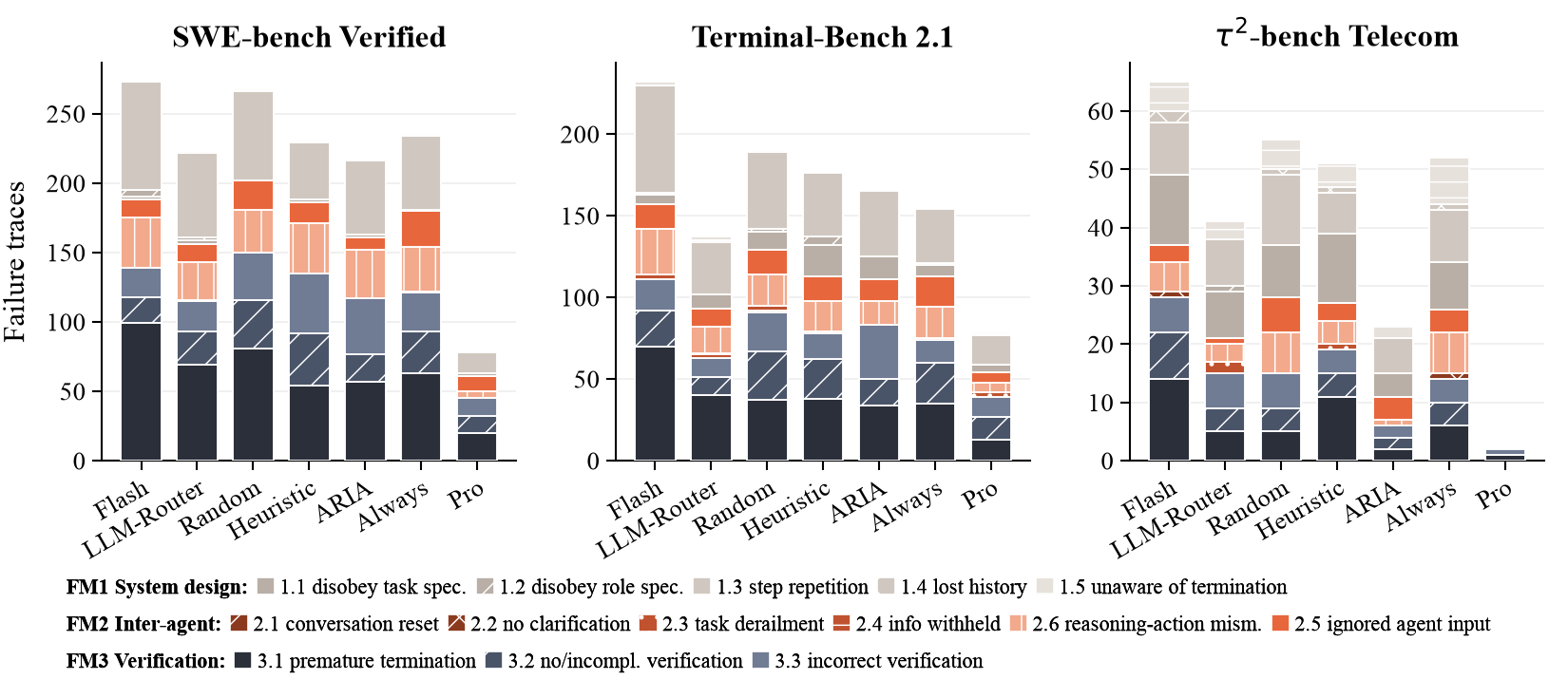}
    \caption{Secondary failure modes across baselines and benchmarks grouped into system-design, inter-agent misalignment, and verification failures, showing different residual failure profiles.}
    \label{fig:mast_secondary}
\end{figure}

\paragraph{Secondary failure modes.}
The main-text analysis focuses on the mutually exclusive primary failure mode
to characterize the dominant reason each unsuccessful trajectory breaks down.
Figure~\ref{fig:mast_secondary} complements this view with all additional MAST
modes observed in the same failed trajectories. Because these presence labels
are non-exclusive, a single trajectory can contribute to multiple categories,
revealing failure mechanisms that co-occur rather than determining only which
one is judged primary.

The broader profiles reinforce the distinction highlighted in Section~5.4.
On SWE-bench, verification-related failures remain common across conditions,
but externally allocated consultation also introduces substantial
inter-agent failure mass, particularly ignored agent input. Thus, improved
consultation placement can remove one bottleneck while exposing another:
relevant guidance may enter the trajectory without being incorporated by the
executor. On Terminal-Bench, the secondary profiles remain more mixed,
consistent with the finding that \app{} and \textit{Heuristic} can achieve
similar aggregate success while failing on different tasks and for different
reasons. On $\tau^2$-bench, inter-agent failure modes are comparatively less
prominent, matching the much lower ignored-input rates discussed in the main
text. The secondary labels therefore provide a more complete view of residual
system behavior without changing the primary-failure comparisons reported in
Section~5.4.

\end{document}